\documentclass[aps,twocolumn,superscriptaddress,amsmath,longbibliography]{revtex4-2}  
\usepackage{graphicx}  
\usepackage{dcolumn}   
\usepackage{bm}        
\usepackage{amsmath,amsfonts,amssymb}   
\usepackage[colorlinks]{hyperref}
\usepackage{floatrow}
\usepackage[caption=false]{subfig}
\usepackage{xcolor}

\newtheorem{theorem}{Theorem}

\newcommand{\CC}{\mathbb{C}}
\newcommand{\RR}{\mathbb{R}}

\newcommand{\ConditionallyIndependent}[3]{#1 \perp\kern-5pt \perp #2 \mid #3}

\newcommand{\set}[1]{\lbrace #1 \rbrace}
\newcommand{\abs}[1]{\lvert #1 \rvert}
\newcommand{\norm}[1]{\lVert #1 \rVert}

\newcommand{\ket}[1]{\lvert #1 \rangle}

\newcommand{\infint}{\int_{-\omega_c}^{\omega_c}}

\newcommand{\so}{S(\omega)}

\DeclareMathOperator{\EE}{\mathbb{E}}
\DeclareMathOperator{\Var}{Var}

\DeclareMathOperator{\sinc}{sinc}

\newcommand{\beq}{\begin{equation}}
\newcommand{\eeq}{\end{equation}}

\begin{document}

\title{Random Pulse Sequences for Qubit Noise Spectroscopy}
\author{Kaixin Huang}
\affiliation{Joint Quantum Institute (JQI), NIST/University of Maryland, College Park, MD 20742, USA}
\affiliation{Joint Center for Quantum Information and Computer Science (QuICS),
NIST/University of Maryland, College Park, MD 20742, USA}
\author{Demitry Farfurnik}
\affiliation{Department of Electrical and Computer Engineering, and Department of Physics, North Carolina State University, Raleigh, NC, 27695, USA}
\author{Alireza Seif}
\affiliation{Joint Quantum Institute (JQI), NIST/University of Maryland, College Park, MD 20742, USA}
\affiliation{Joint Center for Quantum Information and Computer Science (QuICS),
NIST/University of Maryland, College Park, MD 20742, USA}
\affiliation{Pritzker School of Molecular Engineering, University of Chicago, Chicago, IL 60637, USA}
\author{Mohammad Hafezi}
\affiliation{Joint Quantum Institute (JQI), NIST/University of Maryland, College Park, MD 20742, USA}
\affiliation{Joint Center for Quantum Information and Computer Science (QuICS),
NIST/University of Maryland, College Park, MD 20742, USA}
\author{Yi-Kai Liu}
\affiliation{Joint Center for Quantum Information and Computer Science (QuICS),
NIST/University of Maryland, College Park, MD 20742, USA}
\affiliation{Applied and Computational Mathematics Division, National Institute of Standards and Technology (NIST), Gaithersburg, MD 20899, USA}

\noaffiliation

\date{\today}

\begin{abstract}
Qubit noise spectroscopy is an important tool for the experimental investigation of open quantum systems. However, conventional techniques for noise spectroscopy are time-consuming, because they require measurements of the noise spectral density at many different frequencies. Here we describe an alternative approach to noise spectroscopy, which requires fewer resources, and relies on direct measurement of arbitrary linear functionals of the noise spectral density. This method uses random pulse sequences with carefully-controlled correlations, which are chosen using algorithms for phase retrieval. These measurements allow us to  reconstruct sparse noise spectra via compressed sensing. Our simulations of the performance of the random pulse sequences on a realistic physical system, self-assembled quantum dots, reveal a speedup of an order of magnitude in extracting the noise spectrum, compared to conventional dynamical decoupling approaches. 
\end{abstract}

\maketitle

\section{\label{intro}Introduction}

Noise spectroscopy is an essential tool for understanding the behavior of a quantum system coupled to an environment. It plays an important role in the experimental investigation of quantum computation and quantum sensing, in physical systems such as superconducting qubits, semiconductor quantum dots, and nitrogen-vacancy centers in diamond \cite{yan2012spectroscopy, dial2013charge, muhonen2014storing, reinhard2012tuning, bar2012suppression, romach2015spectroscopy, degen2017quantum}. Typically, noise spectroscopy consists of estimating the noise spectral density, $S(\omega)$, at many different frequencies $\omega$, using techniques such as relaxometry or dynamical decoupling \cite{multiseq, alvarez2011measuring, young2012qubits, norris2016qubit, paz-silva2017multiqubit, ferrie2018bayesian}. 

Dynamical decoupling (DD) pulse sequences have been studied for decades in the field of nuclear magnetic resonance (NMR) to reduce the dephasing of spin ensembles \cite{haeberlen1976advances, vandersypen2005nmr},  and later implemented in various quantum systems for noise spectroscopy \cite{bylander2011noise,wang2021intrinsic,bar2012suppression,malinowski2017notch,chan2018assessment,multiseq}. The rotation $\pi$-pulses incorporated in these sequences shape the filter function that probes the qubit's environment in the frequency domain. However, this requires the application of many pulse sequences to learn $S(\omega)$ across the whole frequency domain, and is thus quite time-consuming. This problem is exacerbated when one performs multi-qubit noise spectroscopy on systems with many qubits, which are relevant to many promising applications of quantum technologies \cite{paz-silva2017multiqubit, Preskill2018quantumcomputingin}.

In this work, we develop a different approach to performing noise spectroscopy, which requires fewer resources, and relies on direct measurement of arbitrary linear functionals of the noise spectral density $S(\omega)$. 
This can be used to estimate physically-relevant properties of $S(\omega)$, \textit{without} needing to completely characterize $S(\omega)$. Some properties of $S(\omega)$ that can be measured in this way include: the total noise strength, which can be used to detect changes in the environment of the qubit in real time; inner products between $S(\omega)$ and sinusoidal functions, which are useful for compressed sensing of $S(\omega)$ (described later in this paper).

Our approach applies $\pi-$pulses at random timings with carefully-chosen correlations (i.e., random pulse sequences). 
These random pulse sequences are generated using finite impulse response (FIR) filters, in a way that is simple enough to be implemented in many experimental setups. The design of these FIR filters relies on computational methods for \textit{phase retrieval}, which is the task of estimating a function, given measurements of the (squared) absolute value of its Fourier transform \cite{jaganathan2016phase}. Algorithms for phase retrieval have a long history of use for image reconstruction in astronomy, crystallography and other fields. Our use of phase retrieval, for sensing in the time and frequency domains, is relatively unusual.

The first main result of this paper is to show that phase retrieval techniques can be used to design FIR filters for measuring any desired linear functional of $S(\omega)$, subject to some mild admissibility conditions. In addition, we show that the statistical fluctuations in these measurements, due to the random choice of the pulse sequence, become small as one increases the number of pulses $M$. More precisely, we show that the magnitude of these fluctuations is of order $1/\sqrt{M}$, relative to the expectation value of the measurement.

These results can be compared with other recent works on generating or simulating noise with prescribed time-correlations that are much slower than the system dynamics \cite{serinaldi2017betabit, serinaldi2017general, schultz2021schwarma, murphy2022universal}. Our goal in this paper is different, however: we use random pulse sequences to measure properties of the noise generated by an unknown environment, rather than to simulate or model a source of noise that has already been characterized by some other kind of measurement. 

The second contribution of this paper is to demonstrate efficient reconstruction of sparse noise spectra $S(\omega)$, using a combination of the random pulse sequences described above, and techniques from compressed sensing \cite{candes2008introduction,candes2011probabilistic}. 
This method is reminiscent of compressed sensing techniques used in NMR, though the domain where we apply these techiques (noise spectroscopy) is quite different \cite{https://doi.org/10.48550/arxiv.2109.13298, bostock2017compressed}.
This method requires measurements of only $O(s\log n)$ linear functionals of $S(\omega)$, where $s$ is the sparsity and $n$ is the number of grid points in the frequency domain. 

For a realistic physical system that has a sparse noise spectrum, self-assembled quantum dots \cite{stockill2016quantum, farfurnik2021all}, numerical simulations show that this method can achieve an order of magnitude speedup, compared to conventional dynamical decoupling sequences.
In addition, proof-of-concept demonstrations on commercially-available quantum information processors, using superconducting qubits and trapped ions, and synthetically generated noise with correlations that are much slower than the system dynamics \cite{schultz2021schwarma, murphy2022universal}, show the practical feasibility of this approach.

\section{noise model}
Let us consider a single qubit (``the system") coupled to a 
classical bath that leads to pure dephasing of the qubit. The general Hamiltonian can be written as
\beq
\hat{H}(t)=\hat{H}_0+\hat{H}_V(t)=(\Omega+V(t))\sigma_z,
\eeq
where $\hat{H}_0=\Omega\sigma_z$ is the system Hamiltonian and $\hat{H}_V(t)$ is  the Hamiltonian associated with a  stochastic process $V(t)$ that describes the noise caused by the bath.  For example, $V(t)$ can represent a classical fluctuating variable, such as a magnetic field. For simplicity, here we assume that $V(t)$ is a Gaussian process with zero mean value 
\beq
\langle V(t) \rangle_V=0,
\eeq
where $\langle ...\rangle_V$ stands for the average with respect to the ensemble of $V(t)$.  The Gaussian process  is determined by the auto-correlation
\beq
\langle V(t)V(t') \rangle_V=g(t-t').
\eeq
In the frequency domain, the spectral density can be defined by the Fourier transform of the  auto-correlation,
\beq
\so=\int_{-\infty}^{+\infty} e^{-i \omega t}g(t) dt.
\eeq
The dynamics of a system coupled to a Gaussian bath can be entirely determined by the spectrum, $\so$ \cite{theoryofoqs,qds}. 

In this work, we make an additional assumption that the noise spectrum vanishes at  frequencies larger than a cutoff frequency $\omega_c$, that is, $S(\omega)=0$ when $|\omega| > \omega_c$ \cite{viola1999dynamical, khodjasteh2011limits}. This frequency $\omega_c$ is also referred as `ultraviolet cutoff' for Ohmic spectra and many $1/f$ spectra \cite{cywinski2008howto, ithier2005decoherence}. We assume that the experimentalist has access to sufficiently fast controls, to probe the noise spectrum up to this cutoff frequency.

The methods for noise spectroscopy described in this paper can also be extended to characterize quantum environments, such as bosonic baths \cite{leggett1987dynamics}. This involves a technical complication, as the noise spectrum $S(\omega)$ is no longer an even function. However, when the bath is at thermal equilibrium, the asymmetry of $S(\omega)$ has a simple structure that is determined by the temperature of the bath. If this temperature is known, then $S(\omega)$ can be fully characterized (see Appendix \ref{sec-spin-boson-model}).

\section{Protocols for noise spectroscopy}\label{dyd}
\begin{figure}
\begin{centering}
\sidesubfloat[]{\label{pulse1}%
  \includegraphics[width=0.8\textwidth]{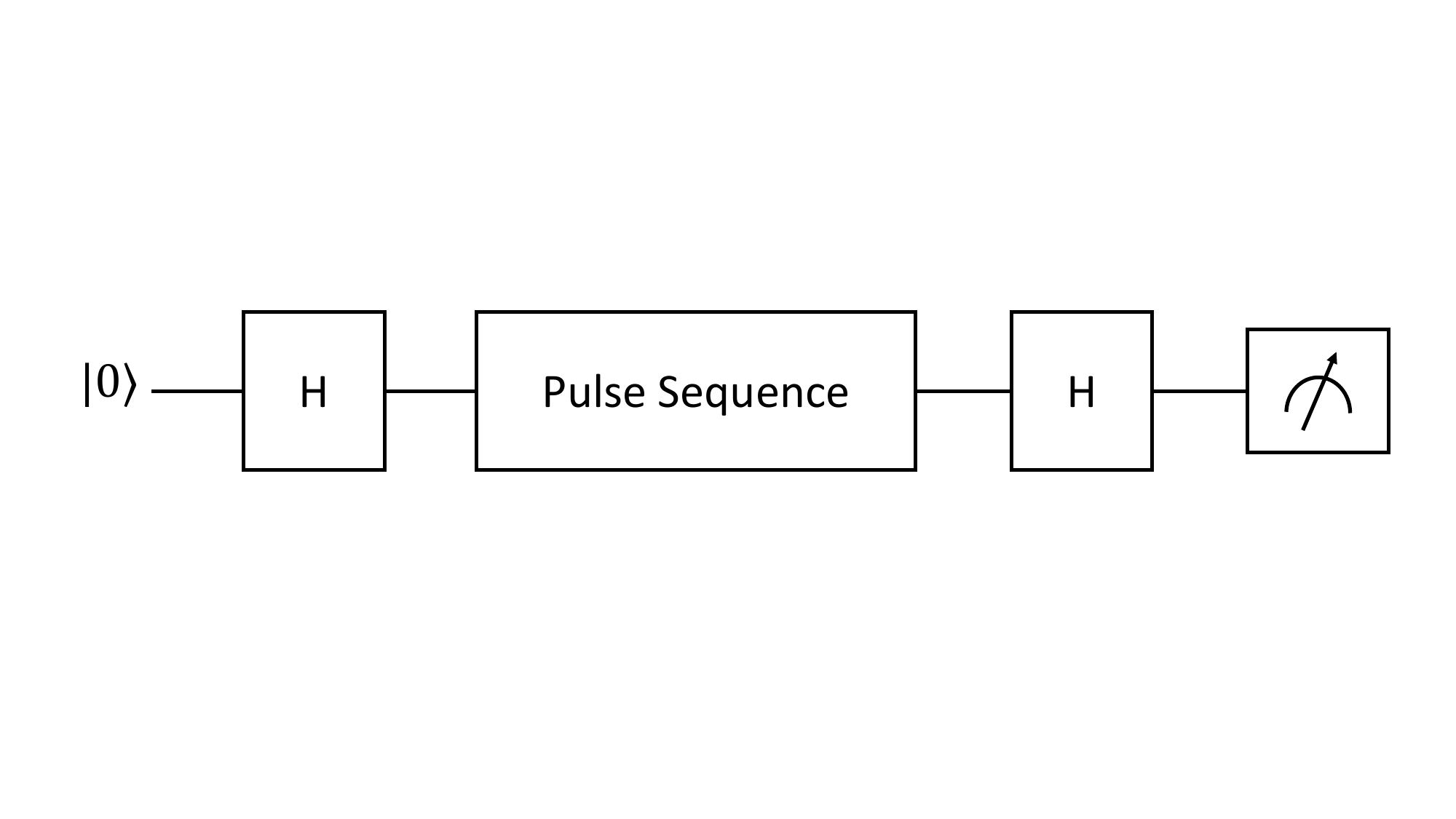}%
}
\par\end{centering}
\begin{centering}
\sidesubfloat[]{\label{pulse2}%
  \includegraphics[width=0.8\textwidth]{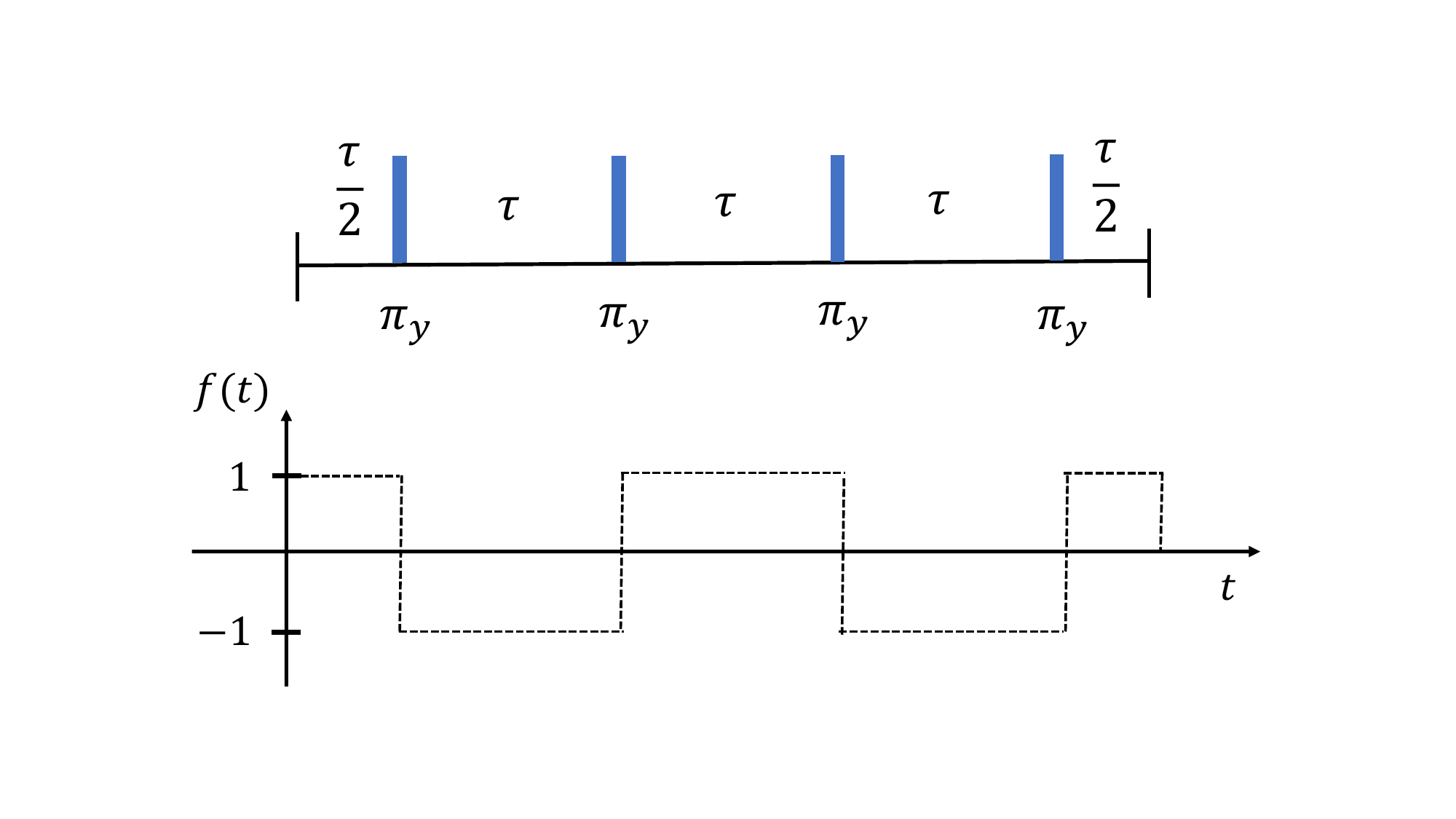}%
}
\par\end{centering}
\begin{centering}
\sidesubfloat[]{\label{rpsillus-c}%
  \includegraphics[width=0.8\textwidth]{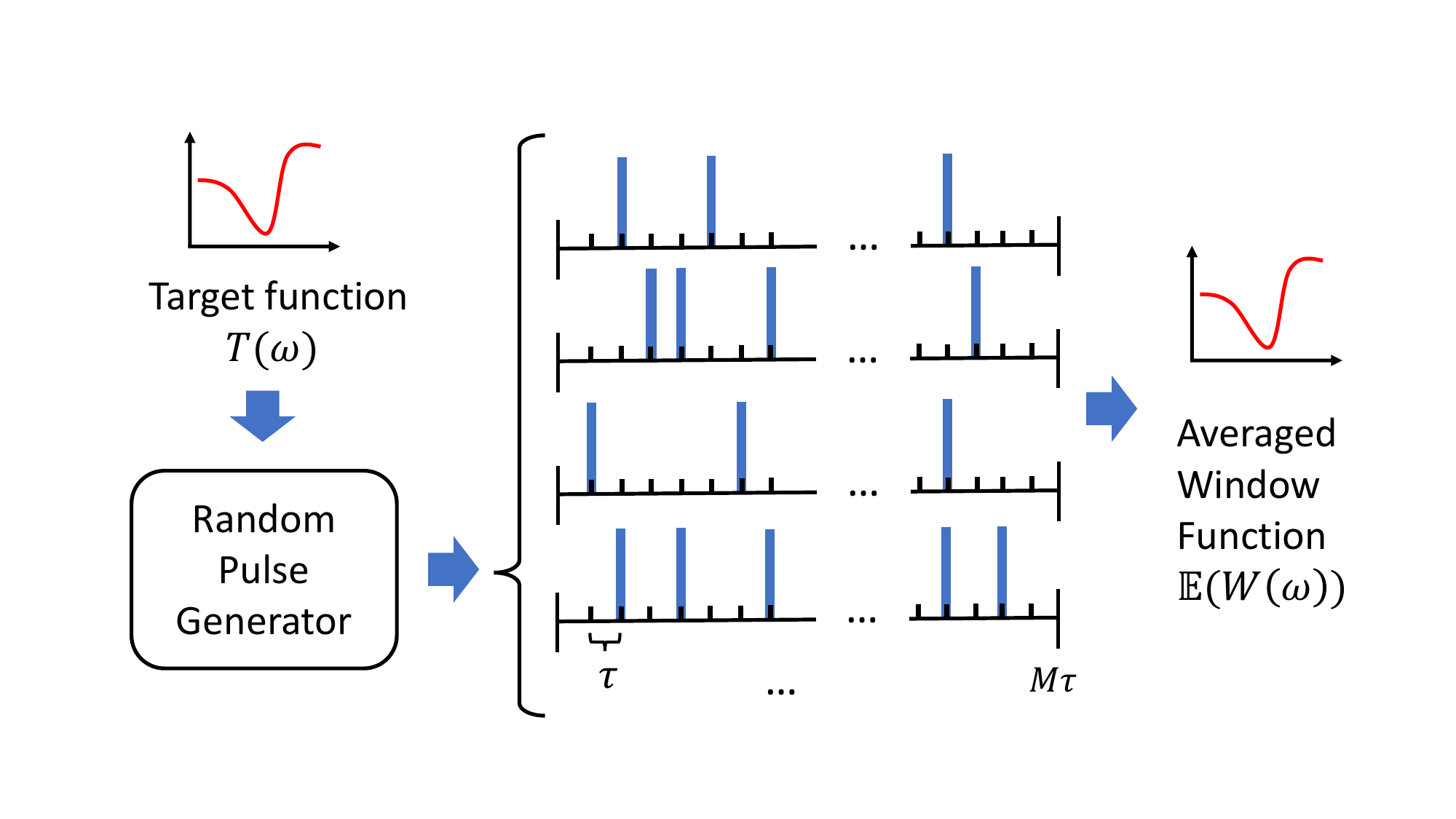}%
}
\par\end{centering}
\caption{(a). A single-qubit noise spectroscopy experiment. The qubit is initialized in the $|+\rangle$ state by a Hadamard gate. Then, the qubit evolves under the application of the pulse sequence. Finally, a measurement is performed in the $\sigma_z$ basis after another Hadamard gate. (b). Top: An illustration of the CPMG pulse sequence with $M=4$ instantaneous $\pi$-pulses (blue rectangles) along the $y$ axis. The time interval between each two pulses is $\tau=T/M$. The qubit freely evolves for a time of $\tau/2$ before the first pulse and after the last one. Bottom: The corresponding filter function, $f(t)$, flips the sign every time a $\pi$-pulse is applied. (c) Illustration of the random pulse sequence method. Given a target function, $T(\omega)$, a random pulse generator can produce multiple pulse sequences. The blue spikes represent instantaneous $\pi$-pulses. The expectation value of the resulting window function, $\EE(W(\omega))$, approximates $T(\omega)$. As a result, one can measure the linear functional $\int S(\omega)T(\omega)d\omega$ by averaging the outcome of each individual pulse sequence.}
\end{figure}

As illustrated by Fig. \ref{pulse1}, a general protocol for noise spectroscopy goes as follows: (1) Prepare the system qubit in the $|+\rangle=\frac{1}{\sqrt{2}}(|0\rangle+|1\rangle)$ state using a Hadamard gate. (2) Apply a sequence of $\pi$ pulses, of 
{total} time duration $T$. (3) Rotate the qubit back with a Hadamard gate and  measure its state in the $\sigma_z$ basis. (4) Repeat (1)-(3)  many times and estimate the probability, $P_{0}(T)$, to obtain the qubit in the $|0\rangle$ state.

For a stochastic bath, $P_{0}(T)$ yields an exponential decay, $e^{-\chi(T)}$, which only depends on $\so$ and the pulse sequence \cite{PhysRevLett.113.250501, PhysRevA.95.022121, cywinski2008howto},
\begin{equation}
\label{coherence}
\begin{aligned}
P_{0}(T) &= \tfrac{1}{2}(1+e^{-\chi(T)}),\\
\chi(T)&=\int_{-\infty}^{\infty}\frac{d\omega}{2\pi} \so |\tilde{f}(\omega)|^2 =\int_{-\infty}^{\infty} \frac{d\omega}{2\pi} \so W(\omega),
\end{aligned}
\end{equation} where $f(t)$ is the filter function corresponding to the pulse sequence (see Fig.~\ref{pulse2}), $\tilde{f}(\omega)$ is the Fourier transform of the filter function,  and the window function, $W(\omega)$, is defined as $W(\omega) = |\tilde{f}(\omega)|^2$. Note that $f(t)$, $\tilde{f}(\omega)$ and $W(\omega)$ all depend implicitly on the total evolution time $T$.

\subsection{Dynamical decoupling}
Here we review the  Carr-Purcell-Meiboom-Gill (CPMG) pulse sequence  as the typical DD method used for noise spectroscopy \cite{haeberlen1976advances,bylander2011noise,wang2021intrinsic,bar2012suppression,malinowski2017notch,chan2018assessment,yuge2011measurement,farfurnik2021all}.
The CPMG pulse sequence (illustrated by Fig. \ref{pulse2}) consists of $M$ rotation $\pi$-pulses applied with time period $\tau$ ($\tau=T/M$). The window function of this sequence equals
\beq
\label{dd}
\begin{aligned}
W(\omega)=\begin{cases} \frac{32}{\omega^2}\sin^4(\tfrac{\omega \tau }{4})\sin^2(\tfrac{\omega M\tau }{2})/\cos^2(\tfrac{\omega \tau }{2}), \text{for even $M$}\\
\frac{32}{\omega^2}\sin^4(\tfrac{\omega \tau }{4})\cos^2(\tfrac{\omega M\tau }{2})/\cos^2(\tfrac{\omega \tau }{2}), \text{for odd $M$}.
\end{cases}
\end{aligned}
\eeq
 When $M$ is large, $W(\omega)$ can be approximated as a Dirac $\delta-$function located at $\omega=\tfrac{\pi}{2} \tau$. As a result, the application of a CPMG sequence mainly probes the noise at a single frequency of $\tfrac{\pi}{2} \tau$. However, the exact window functions in Eq.~(\ref{dd}) deviate from Dirac $\delta-$functions and include higher harmonics. A commonly used technique to deal with deviations is the Alvarez-Suter method \cite{alvarez2011measuring, szankowski2018accuracy, szankowski2019transition}, which applies a comb approximation to the window function when $M$ approaches infinity. In this work, we apply a
deconvolution procedure to correct for the deviations without any assumption on $M$, details described in Appendix \ref{sec-cpmg-2nd-order-corrections}. 

To fully decompose the spectrum then requires sweeping the free evolution time, $\tau$ (or the number of pulses, $M$). However, such probing across the whole frequency range involves numerous measurements, which are not always necessary for extracting the significant features of the noise.

\subsection{Random pulse sequences}
\label{sec-random-pulse-sequences}
We propose an alternative method for noise spectroscopy based on random pulse sequences. Rather than probing single frequencies like the CPMG method, our approach generates window functions, $W(\omega)$, that approximate a desired spectral shape $T(\omega)$. 
The general idea is to design a \textit{random} pulse generator to produce a group of sequences, such that the expectation value of their window function, $\EE(W(\omega))$, can approximate the desired target function, $T(\omega)$, as illustrated in Fig. \ref{rpsillus-c}. 
As a result, we can directly estimate the linear functional, $I=\int \so T(\omega)d\omega$. This procedure is analogous to the generation of a stationary sequence \cite{rosenblatt2012stationary}, adapted for noise spectroscopy.

The random pulse sequences are generated in the following way. The total experiment time, $T$, is divided into $M$ equal segments, such that $T=M\tau$.  Rotation $\pi$-pulses are applied only at the end of particular segments determined by a random pulse generator. Specifically, we generate a vector of random variables, $\vec{U} = (U_1,\ldots,U_M) \in \set{1,-1}^M$, 
{and a corresponding random pulse sequence}, such that: 
\begin{enumerate}
\item $U_i$ represents the value of the filter function, $f(t)$, in the time segment $t \in [(i-1)\tau,i\tau]$. A $\pi$-pulse is applied at time $i\tau$ if and only if $U_i \neq U_{i+1}$. 
\item The expectation value of any random variable is zero, i.e., $\EE(U_i)=0$. 
\item The covariance of two random variables, $U_i, U_j$, should only depend on the distance $\abs{j-i}$, i.e., it has the form $\EE(U_iU_{i+j})=R(j)$ (for $j\geq 1$). 
\end{enumerate}

Random variables $U_i$ satisfying these properties can be constructed by generating a sequence of independent Gaussian random variables $N_0,N_1,N_2,\ldots$, applying a finite impulse response (FIR) filter with suitably chosen coefficients $(a_0, a_1, \ldots, a_{\lambda-1})$, and then applying the sign function to obtain:
\begin{equation}\label{eqn-Ui-aj}
U_i=\text{sign}(\sum_{j=0}^{\lambda-1} a_j N_{i+j}), \quad i=0,1,2,\ldots. 
\end{equation}
In some cases, time-varying FIR filters may be used for improved computational efficiency (see Appendix \ref{sec-alternative-generator}).

The above construction is related to the phase retrieval problem, in the following way. An elementary calculation (see Appendix \ref{sec-generating-random-pulse-sequences-1}) shows that
\begin{equation}
R(k) =\frac{2}{\pi}\arcsin{(\sum_{i=0}^{\lambda-1-k} a_i a_{i+k})},
\end{equation}
i.e., the correlations $R(k)$ in the random pulse sequence are related to the autocorrelation function of the filter coefficients $a_i$. One can interpret the $a_i$ as Fourier coefficients of some periodic function on the real line:
\begin{equation}
b(m) =\sum_{j=0}^{\lambda-1} a_j e^{-j\frac{2\pi im}{2\lambda-1}}.
\end{equation}
Then the $R(k)$ can be rewritten in terms of $|b(m)|^2$, i.e., they encode information about the magnitudes, but not the complex phases, of the function $b(m)$. This is analogous to a phase retrieval problem.

Now, given some desired correlations $\tilde{R}(k)$, let our goal be to choose filter coefficients $a_i$ that will generate random variables $U_i$ whose correlations $R(k)$ approximately match the $\tilde{R}(k)$. We will do this using phase retrieval techniques, in the following way. We interpret the $\tilde{R}(k)$ as ``noisy'' measurements of $|b(m)|^2$, where the ``noise'' is due to the fact that the desired correlations $\tilde{R}(k)$ may not be exactly realizable by a stochastic process of the form (\ref{eqn-Ui-aj}). We then use algorithms for phase retrieval to learn a function $b(m)$ such that $|b(m)|^2$ is approximately compatible with the desired correlations $\tilde{R}(k)$. From $b(m)$ we finally obtain the corresponding filter coefficients $a_i$. 

Assuming that the $\tilde{R}(k)$ satisfy some mild admissibility conditions, the above phase retrieval problem is guaranteed to have a feasible solution, due to the Fej\'er-Riesz theorem \cite{fejer-riesz, fejer1916trigonometrische}. In practice, solutions can be found using gradient descent methods, or the Gerchberg-Saxton (GS) algorithm \cite{jaganathan2016phase}. See Appendix \ref{sec-generating-random-pulse-sequences-2} for details.

\subsection{Measuring Linear Functionals of $S(\omega)$}

We now combine all of the above pieces, in order to show, given a target function $T(\omega)$, how to construct the corresponding random pulse sequence that estimates the linear functional $I=\int \so T(\omega)d\omega$. 

The random pulse sequence produces a certain window function, $W(\omega)$,  that probes the noise. The expectation value of the window function over all the possible realizations of $\vec{U}$ yields
\begin{equation}
\small
\label{windowrps}
\EE(W(\omega)) = M\tau^2\sinc^2(\tfrac{\omega\tau}{2})[1+2\sum_{k=1}^{\lambda}R(k)\cos{(k\omega\tau)}(1-\tfrac{k}{M})],
\end{equation}
where we define $\sinc x = \tfrac{\sin x}{x}$ and $\lambda$ is the cutoff distance of the correlation between random variables, i.e., $R(k)=0$ whenever $k>\lambda$. 

Our goal is to ensure that $\EE(W(\omega))$ approximates some prescribed target function $T(\omega)$. Note that the cosine functions $\{\cos(k\omega\tau)\}$ form an almost complete basis (the zeroth term excluded) in the region $[-\tfrac{\pi}{\tau}, \tfrac{\pi}{\tau}]$. Thus, we take the following approach: we match the time interval between segments and the cutoff frequency of the noise (i.e., setting $\tau=\tfrac{\pi}{\omega_c}$), and we adjust the random pulse generator (i.e., optimizing the filter coefficients $(a_0, a_1, \ldots, a_{\lambda-1})$, using techniques from phase retrieval, as explained in Appendix \ref{sec-generating-random-pulse-sequences}), so that 
\beq 
\label{rk}
R(k) =\tfrac{M}{\pi(M-k)} \int_{-\omega_c}^{\omega_c}\frac{cT(\omega)\cos(k\omega\tau)}{\sinc^2(\tfrac{\omega\tau}{2})}d\omega \quad (k=1,\ldots,\lambda),
\eeq 
which implies
\beq 
\label{ew}
\EE(W(\omega)) \xrightarrow{\lambda\rightarrow \infty} M\tau^2[cT(\omega)+(1-cT_0)\sinc^2(\tfrac{\omega\tau}{2})].
\eeq 

In Eqs.~(\ref{rk}) and (\ref{ew}), $c$ is a parameter that ``rescales'' the target function $T(\omega)$, in order to ensure that it can be approximated by a window function $W(\omega)$ that is generated by a stochastic process of the form (\ref{eqn-Ui-aj}). We show that a suitable value of the parameter $c$ always exists, and describe a heuristic way of setting the parameter $c$ (see Appendix \ref{sec-generating-random-pulse-sequences-2}). Essentially, we want to choose $c$ as large as possible, subject to certain constraints (which can be easily checked). $T_0$ is a constant term depending on $T(\omega)$,
\beq 
\label{t0}
T_0 = \tfrac{1}{\omega_c} \int_{-\omega_c}^{\omega_c}\frac{T(\omega)}{\sinc^2(\tfrac{\omega\tau}{2})}d\omega.
\eeq 

Here, we have constructed the random pulse sequence $\vec{U}$ so that $R(k)$ is proportional to the $k$-th coefficient of the Fourier series representation  of $T(\omega)/\sinc^2(\tfrac{\omega\tau}{2})$ (Eq.~(\ref{rk})). As such, Eq.~(\ref{ew}) is a good approximation for a finite $\lambda$, if $R(k)$ converges to 0 as $k \rightarrow \infty$. By plugging Eq.~(\ref{ew}) into Eq.~(\ref{coherence}), the expectation value of the decay exponent yields
\begin{equation}
\label{chirps}
\small
    \EE(\chi)=\frac{M\tau^2}{2\pi}\int_{-\omega_c}^{\omega_c}[cT(\omega)S(\omega)d\omega+(1-cT_0)\sinc^2(\tfrac{\omega\tau}{2})S(\omega)d\omega].
\end{equation}
Note that $\EE(\chi)$ grows linearly with the total evolution time $T = M\tau$, as expected.

From Eq.~(\ref{chirps}), we can extract the desired functional, $I=\int \so T(\omega)d\omega$, by subtracting the second term $\int S(\omega) \sinc^2(\tfrac{\omega\tau}{2})d\omega$ from  $\EE(\chi)$. This term can be estimated by applying an additional series of ``base'' random pulse sequences, for which  $\vec{U}$ contains independent random variables (that is, $R(k)=0$ for all $k>0$). The expectation value of the ``base'' sequence's window function is $\EE(W_{\text{base}}(\omega))=\tfrac{M\tau^2}{2\pi}\sinc^2(\tfrac{\omega\tau}{2})$, and the corresponding decay exponent is $\EE(\chi_{\text{base}})=\tfrac{1}{2\pi}\int \so \EE(W_{\text{base}}(\omega))$. This allows us to subtract the second term in Eq.~(\ref{chirps}). Alternatively, one can set up a Ramsey experiment of time $\tau$ to measure  the second term. The cost for each method depends on experimental details. For example, ideally we want $\chi$ to be of order 1. When $\tau$ is too small, $\chi$ might be too close to 0 for the Ramsey experiment, thus increasing the number of measurements needed to reach certain accuracy. 

\begin{figure}
\begin{centering}
\sidesubfloat[]{\label{fig:FIG4a}%
  \includegraphics[width=0.8\textwidth]{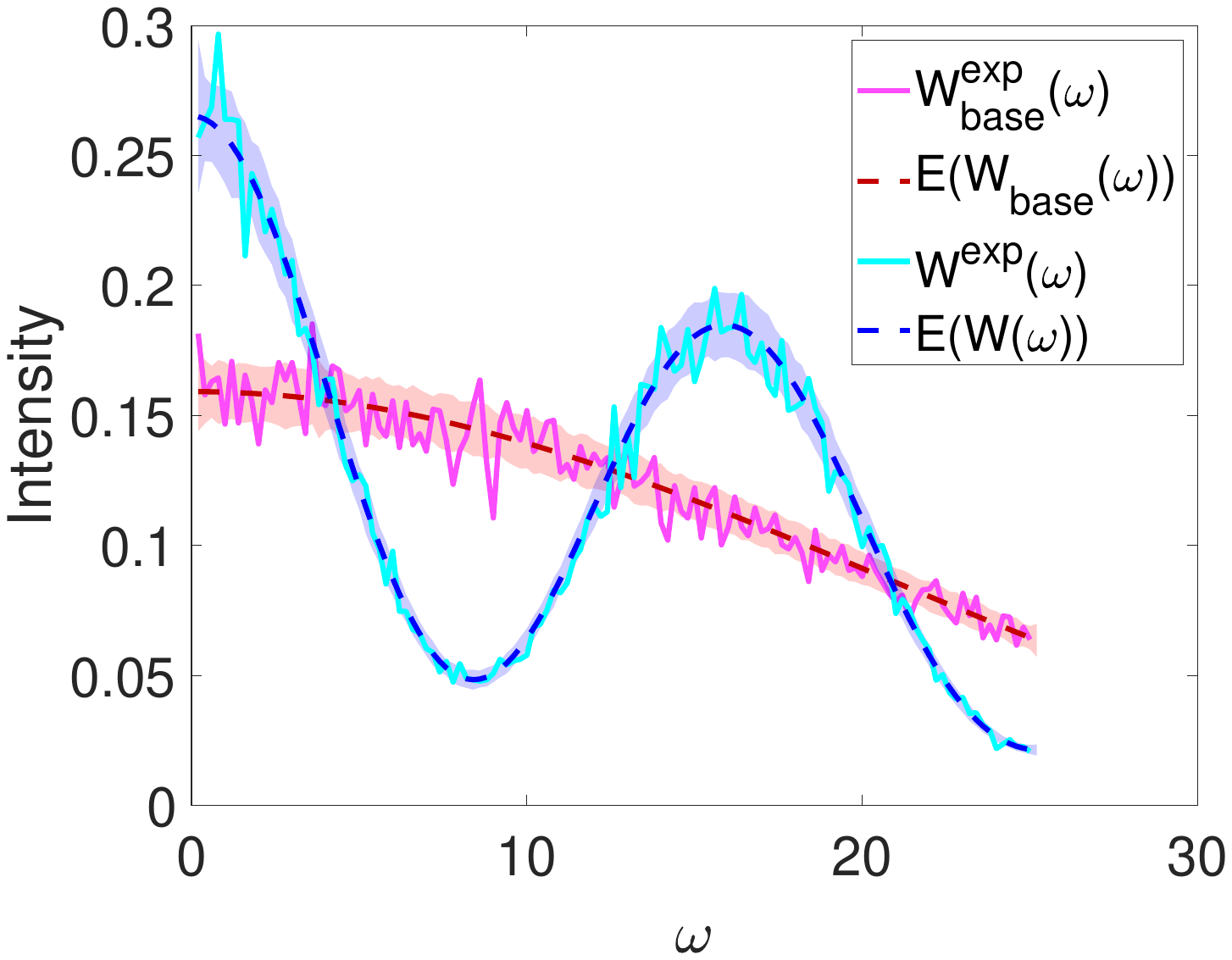}%
}
\par\end{centering}
\begin{centering}
\sidesubfloat[]{\label{fig:FIG4b}%
  \includegraphics[width=0.8\textwidth]{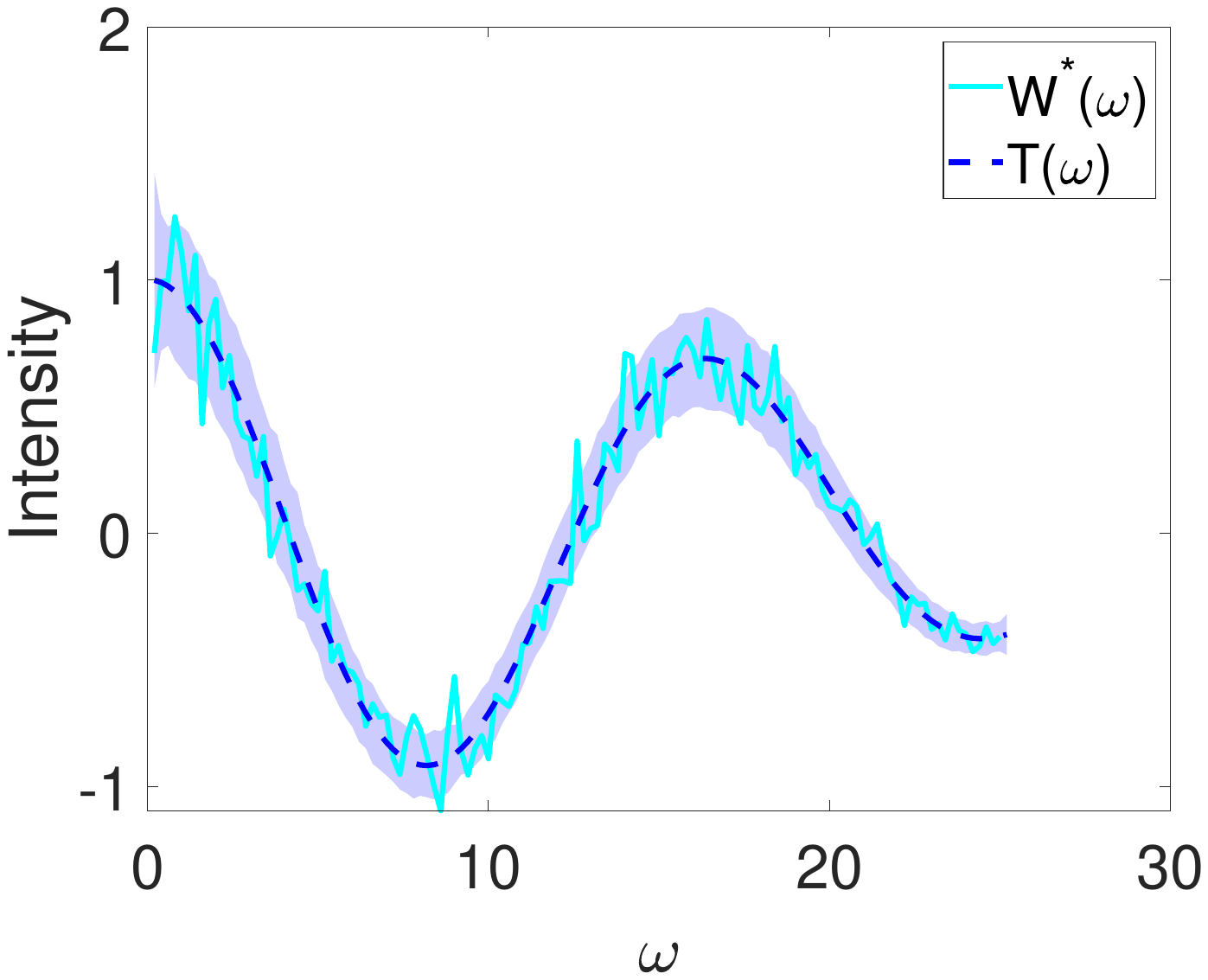}%
}
\par\end{centering}
\caption{
Illustration of the generation of the target function ${T}(\omega)=\sinc^2{({\omega\pi/50})}\cos{(3\omega\pi/25)}$ by two groups of random pulses. In both plots, the dashed lines represent exact expectation values. The solid lines represent simulated results of an experimental realization considering 200 random pulse sequences with 250 segments, each repeated 200 times. The shaded areas represent the $\pm 1$ standard deviation of these results. (a) The dashed blue line (dashed red line) stands for the expectation value of the window function, $\EE(W(\omega))$ (of the base window function, $\EE(W_{\text{base}}(\omega)$)), generated by ${T}(\omega)$, and the solid cyan line (solid magenta line) is one corresponding simulation of the experimental result, $W^{\text{exp}}(\omega)$ ($W_{\text{base}}^{\text{exp}}(\omega)$). (b) The dashed blue line represents ${T}(\omega)$. The solid cyan line stands for the extracted window function of the two averaged window functions in (a), calculated by ${W^*(\omega)} = [W^{\text{exp}}(\omega)-(1-cT_0)W_{\text{base}}^{\text{exp}}(\omega)]/c$. Here $c$ is set to be $1/3$.
}
\label{rpsillus}
\end{figure}

Figure \ref{rpsillus} illustrates the window functions generated by the random pulse protocol for a target function of $T_1(\omega)=\sinc^2{(\tfrac{\omega\tau}{2})}\cos{(3\omega\tau)}$. The value of $\EE(W_{base})$ (dashed magenta line in Fig. \ref{fig:FIG4a}) is subtracted from $\EE(W(\omega))$ (dashed cyan line in Fig. \ref{fig:FIG4a}), to obtain $T(\omega)$ (dashed cyan line in Fig. \ref{fig:FIG4b}).

The experimental estimation of the desired functional, $I^{\text{exp}}$, requires measuring the decay exponent $\chi^{\text{exp}}$, by generating $N_1$ different random pulse sequences with each repeated $N_2$ times, as well as measuring the base decay exponent, $\chi_{\text{base}}^{\text{exp}}$, by generating $N_{\text{base},1}$ different random pulse sequences with each repeated $N_{\text{base},2}$ times (see Appendix \ref{secd3}). 

Note that we are free to choose larger values of $N_1$ and $N_2$ in order to obtain better accuracy; whereas our choices of $T$, $M$ and $\tau$ are more constrained. In particular, $T$ must be chosen so that the decay exponent $\chi$ is of order 1 (so that it can be estimated from a small number of experimental trials); $\tau$ is constrained by the capabilities of the experimental apparatus and its control system; and $M$ must satisfy $T = M\tau$.

For the specific target function $T(\omega)$ in Fig. \ref{rpsillus}, the application of random pulse sequences with $(M,N_1,N_2)=(M,N_{\text{base},1},N_{\text{base},2})=(200,200,50)$ (solid lines in Fig. \ref{rpsillus}) provides a close estimation of the expectation values (dashed lines in Fig. \ref{rpsillus}).

\section{Accuracy of the method}
The accuracy of the experimental estimation of the desired functional, $I^{\text{exp}}$,  depends on the accuracies of the experimentally measured decay exponents, which are also yielded by the method to generate random pulses. This accuracy can be bounded in the following way: 
\begin{equation}
\label{eqn-variance}
\small
\begin{aligned}
    \text{Var}(\chi)\leq\frac{(7\tilde{\lambda}^2 + 10\tilde{\lambda} + 1)M\tau^3}{\pi}||&S(\omega)\sinc^2(\tfrac{\omega\tau}{2})\\-
    & \langle S(\omega)\sinc^2(\tfrac{\omega\tau}{2})\rangle||^2_{L2},
\end{aligned}
\end{equation}
\begin{equation}
\label{accuracy}
\small
\begin{aligned}
    |\chi^{\text{exp}}-\EE(\chi)| 
    = O\bigg(\frac{\sqrt{\text{Var}(\chi)}}{\sqrt{N_1}}
    + \frac{\mathbb{E}(\chi)}{\sqrt{N_1N_2}}\bigg).
\end{aligned}
\end{equation}
Here $\Var(\chi)$ is the variance of the decay rate $\chi$ with respect to the choice of the random pulse sequence. $\tilde{\lambda}$ denotes twice the number of nonzero $R(k)$s, and can be interpreted as an upper bound on the complexity of the correlations among the random pulses. $\langle \cdot \rangle$ denotes averaging over the frequency domain, i.e., $\langle f(\omega) \rangle = \frac{\tau}{2\pi} \int_{-\pi/\tau}^{\pi/\tau} f(\omega) d\omega$. Finally, a bound similar to Eq.~(\ref{accuracy}) holds for $|\chi^{\text{exp}}_{\text{base}}-\EE(\chi_{\text{base}})|$ with $\EE(W_\text{base}(\omega))=M\tau^2\sinc^2(\tfrac{\omega\tau}{2})$. Eqs.~(\ref{eqn-variance}) and (\ref{accuracy}) are derived in Appendix \ref{sec-accuracy-details}. 

One interesting consequence of Eq.~(\ref{eqn-variance}) is that, as the number of pulses $M$ grows large, $\Var(\chi)$ scales roughly linearly with $M$. This implies that the fluctuations in $\chi$ are small (of order $1/\sqrt{M}$) relative to the expectation value of $\chi$: 
\begin{equation}
\frac{\sqrt{\Var(\chi)}}{\EE(\chi)} \propto \frac{\tilde{\lambda}}{\sqrt{M}}.
\end{equation}
Intuitively, the $M$ random pulses behave as if they were independent, up to a correction factor that depends on $\tilde{\lambda}$ (but not on $M$). This intuition is made precise in the proof of Eq.~(\ref{eqn-variance}), by modeling the random variables $U_i$ as a Markov random field with some correlation graph $G$, and observing that $\tilde{\lambda}$ is related to the maximum degree of $G$. (See Appendix \ref{sec-variance-decay-exponent-general-case}.)

Eq.~(\ref{eqn-variance}) also depends on the squared $L_2$ norm of $\so\sinc^2(\tfrac{\omega\tau}{2})$ minus its average value. This can be viewed as a measure of the smoothness of $\so\sinc^2(\tfrac{\omega\tau}{2})$. This is consistent with what one might intuitively expect, when applying a random pulse sequence that samples the noise at many different frequencies simultaneously: when $\so\sinc^2(\tfrac{\omega\tau}{2})$ is roughly constant on all frequencies, the fluctuations in $\chi$ should be small; but when $\so\sinc^2(\tfrac{\omega\tau}{2})$ consists of isolated peaks, the fluctuations in $\chi$ can be large. 

When performing noise spectroscopy on a real physical system, one needs a way to estimate the right hand side of Eq.~(\ref{eqn-variance}), using physical parameters of the environment that may be known \textit{a priori}. This can be done by using Holder's inequality $\norm{A}_{L2}^2 \leq \norm{A}_{L1} \norm{A}_{L\infty}$, and then plugging in upper bounds on the total noise strength, and the maximum noise strength at any single frequency. These bounds can come from knowledge of physical interactions and energy scales in the system.

Finally, Eq.~(\ref{accuracy}) suggests a simple strategy for minimizing the error in the measured decay exponent $\chi^{\text{exp}}$: make $N_1$ large, while keeping $N_2$ as small as possible. One interesting possibility is to set $N_2 = 1$, so that each run of the experiment uses an independently-sampled random pulse sequence. This approach bears some resemblance to fully randomized benchmarking \cite{kwiatkowski2023optimized}. 

Setting $N_2 = 1$ causes difficulties with the statistical estimation of the decay exponent for each random pulse sequence, because there is no averaging to smooth out the shot noise. Nonetheless it is still possible to obtain crude statistical estimates in this situation (see Appendix \ref{sec-acc}). This approach is suitable for experimental setups where the random pulse sequence can be generated in real time, using a field programmable gate array (FPGA) or fast programmable logic. One advantage of this approach is that it removes the need to store the random pulse sequence in computer memory.

\section{Compressed sensing}
\label{sec-compressed-sensing}

\begin{figure}
\begin{centering}
\sidesubfloat[]{\label{fig:cs2}%
  \includegraphics[width=0.8\textwidth]{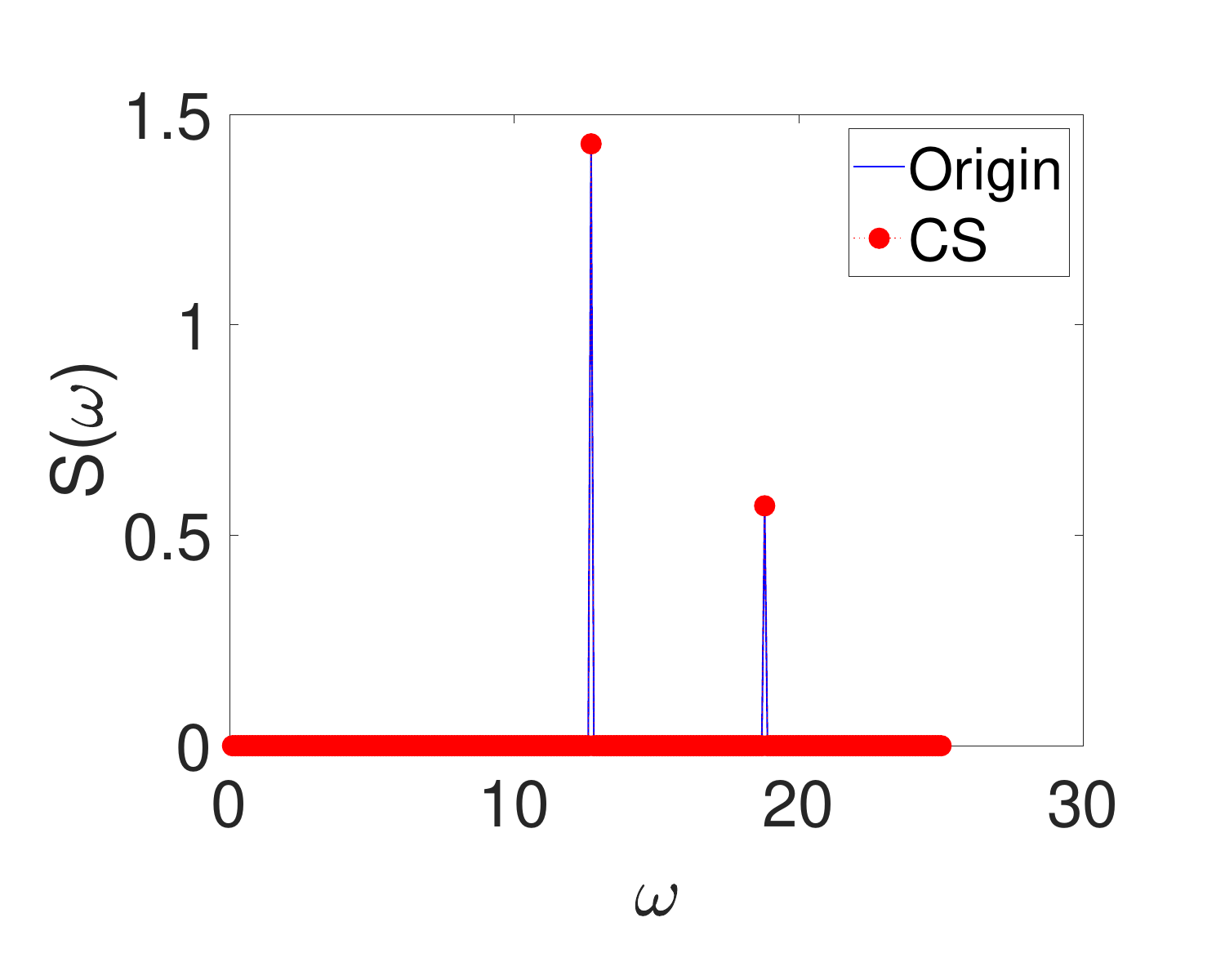}%
}
\par\end{centering}
\begin{centering}
\sidesubfloat[]{\label{fig:csphase}%
  \includegraphics[width=0.8\textwidth]{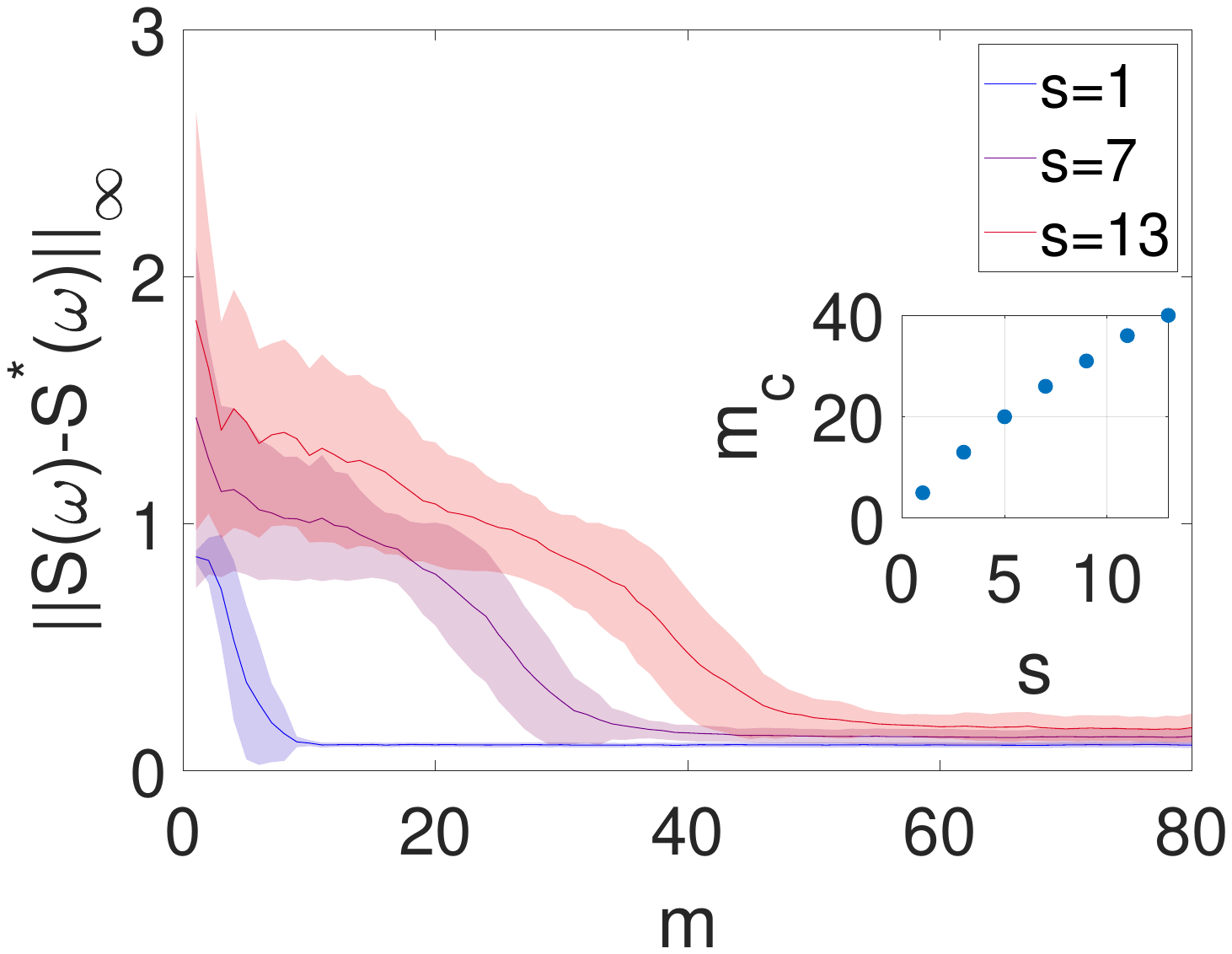}%
}
\par\end{centering}
\caption{ 
(a) A reconstruction of an ideal sparse spectrum using the CS method. The solid blue line represents a 2-sparse spectrum with $N=250$ grid points. The red circles represent the decomposed spectrum using CS based on $m=12$ different Fourier basis functions. For each Fourier basis function, we generate random pulse sequences with $(M, N_1, N_2)=(250, 1000, 50)$. (b) The accuracy of CS ($(M, N_1, N_2)=(250, 1000, 50)$) in reconstructing ideal spectra as a function of the number of Fourier basis functions. Different curves represent different sparsities $s$, considering  200  randomly generated spectra with $N=250$, normalized so that the $L_1$ norm equals $0.1 s$. Each simulation is repeated 100 times and the shaded areas represent the 95\% confidence regime. Inset: The scaling of the critical number of Fourier basis functions, $m_c$, as a function of the sparsity of the spectrum.}
\label{cs}
\end{figure}

\begin{figure}
\includegraphics[width=0.8\textwidth]{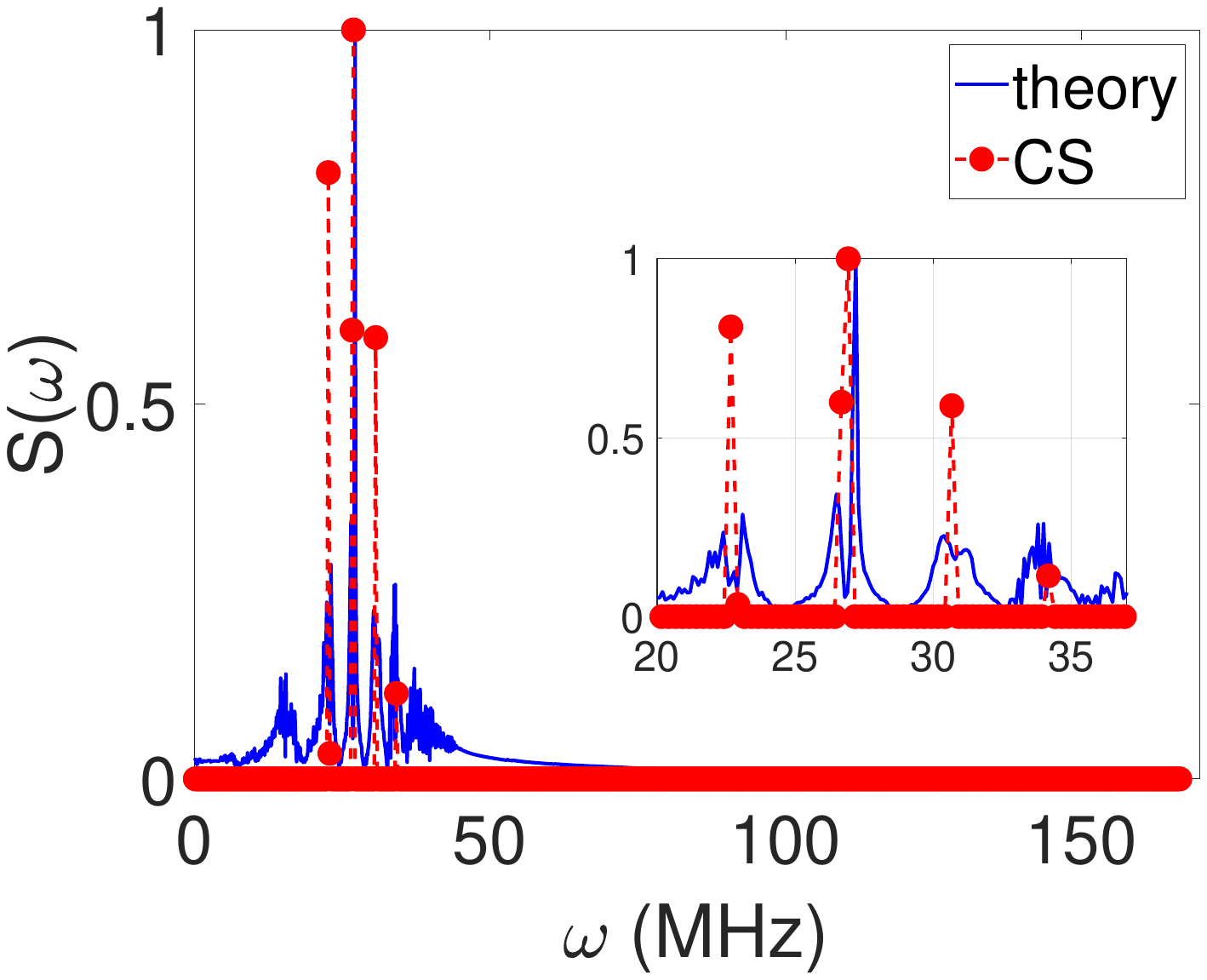}%
\caption{A reconstruction of the noise spectrum of an ensemble of nuclear spins interacting with an InAs/GaAs quantum dot (under an external magnetic field of B = 2 T at the Voigt geometry) using pulse sequences of compressed sensing. The blue solid line represents the theoretically simulated noise spectrum, with the maximum intensity normalized to 1. The red dots represent the simulated reconstructed spectrum considering random pulse sequences with $(M, N_1, N_2)=(200, 2000, 50)$ and $m=40$ different Fourier basis functions. For the CS method, we estimate the spectrum using the LASSO, with $N=667$ grid points and 10-fold cross validation, which successfully identifies the central frequencies of the major peaks.} 
\label{fig:FIG6a}
\end{figure}

\begin{figure}
\includegraphics[width=0.8\textwidth]{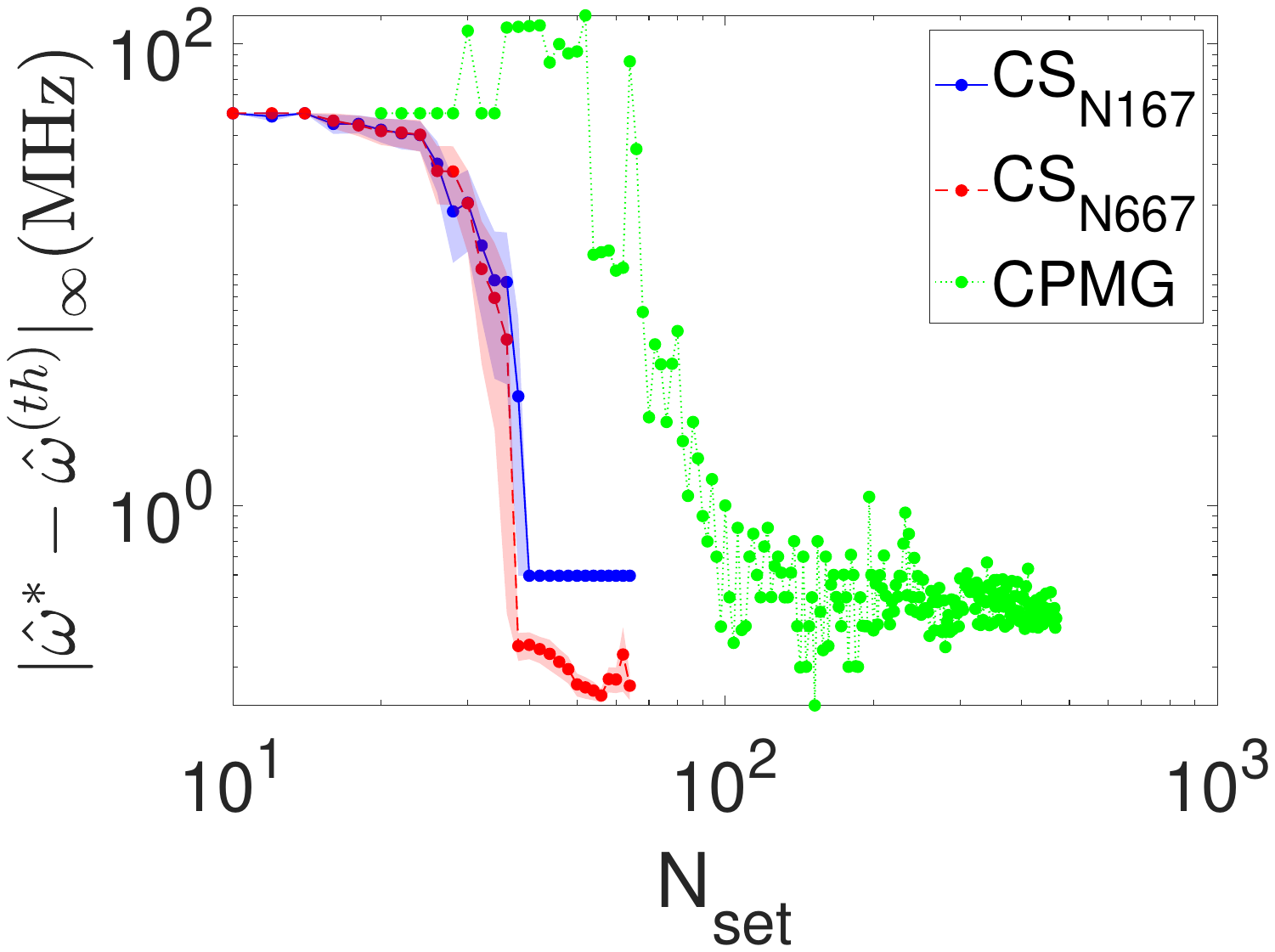}%
\caption{Numerical simulations of the accuracy of reconstructing the central frequencies of the InAs/GaAs noise spectrum as a function of the number of sets of experiments, $N_\text{set}$. The reconstruction accuracy, $|\hat{\omega}^*-\hat{\omega}^{\text{th}}|_{\infty}$, quantifies the deviation between the reconstructed frequencies and the theoretical ones for the three largest peaks. The solid blue line and the dashed red line represent the accuracy of CS with $N=167$ and $N=667$ grid points, respectively. The CS simulations are repeated for 30 times and the shaded areas represents the 95\% confidence regimes. The dotted green line represents the reconstruction accuracy of the noise spectrum using the CPMG sequences. Achieving a certain accuracy by utilizing the CPMG sequences requires an order of magnitude more sets of experiments than by utilizing CS.}
\label{cscompare}
\end{figure}

\subsection{Theory}

A promising application of the random pulse method is the compressed sensing (CS) \cite{candes2008introduction, candes2011probabilistic} of sparse noise spectra.  Let us approximate a noise spectrum by a function on a discrete set of $N$ ``grid points'' in the frequency domain (call this subset $G_N$).
We say the noise spectrum is $s$-sparse if it is nonzero at (at most) $s$ points in $G_N$. As we show below, our method can use $O(s\log N)$ sets of random pulses to fully reconstruct an $s$-sparse spectrum, ideally providing an exponential speedup compared to $O(N)$ sets of experiments required by the CPMG protocol.

The main idea of the CS method is to choose a random set of Fourier basis functions, and then generate random pulse sequences whose window functions approximate these Fourier basis functions, in order to probe the noise spectrum. More precisely, we choose $m$ random Fourier functions with frequencies $k_1,\ldots,k_m$. For each $a \in \set{1,\ldots,m}$, we define target functions $T_{k_a}(\omega) = \cos(k_a\omega\tau)\sinc^2(\tfrac{\omega\tau}{2})$, and we generate random pulse sequences with $R(j)=\tfrac{M}{2(M-k_a)}\delta(j-k_a)$. 
Using equations (\ref{windowrps}) and (\ref{ew}), these random pulse sequences have decay exponents $\chi_{k_a}$ that satisfy 
\begin{equation}
    \label{csbasis}
    \EE(\chi_{k_a}-\chi_{\text{base}})=M\tau^2\int_{-\omega_c}^{\omega_c} S(\omega)\sinc^2(\tfrac{\omega\tau}{2}) \cos(k_a\omega\tau)d\omega.
\end{equation}
We view this as a compressed sensing measurement of the function $S(\omega)\sinc^2(\tfrac{\omega\tau}{2})$, from which we can readily extract the noise spectrum $S(\omega)$. Note that $S(\omega)\sinc^2(\tfrac{\omega\tau}{2})$ has the same sparsity as $S(\omega)$, since  $\sinc^2(\tfrac{\omega\tau}{2})$ varies mildly between $\tfrac{4}{\pi^2}$ and $1$, when $\omega$ is in the interval $[-\tfrac{\pi}{\tau},\tfrac{\pi}{\tau}]$. By the same token, estimating $S(\omega)\sinc^2(\tfrac{\omega\tau}{2})$ is equivalent to estimating $S(\omega)$, up to a small loss of precision.

According to the CS theory \cite{candes2008introduction, candes2011probabilistic}, 
the discretized spectrum $S^*:\: G_N \rightarrow \RR$ (where $G_N$ is the set of grid points) can be recovered by solving a convex optimization problem. Given experimental measurements of $\chi^{\text{exp}}_{k_a}$ ($a=1,\ldots,m$) and $\chi^{\text{exp}}_{\text{base}}$, one can solve:
\begin{equation}
\label{CSconvec-0}
    \begin{aligned}
        &\min_{S^*:\: G_N \rightarrow \RR}
        ||S^*(\omega)\sinc^2(\tfrac{\omega\tau}{2})||_{L_1} ,   \text{ subject to} \\ 
        &\sum_{a=1}^m \Bigl\lvert
        \chi^{\text{exp}}_{k_a}-\chi^{\text{exp}}_{\text{base}} \\
        &- \tfrac{2\omega_c M\tau^2}{N |G_N|} \sum_{\omega\in G_N} S^*(\omega)\sinc^2(\tfrac{\omega\tau}{2}) \cos(k_a\omega\tau)
        \Bigr\rvert^2 \leq \epsilon.\\
    \end{aligned}
\end{equation}
Here, $\epsilon$ is chosen by the experimenter to allow for noise in the measurements of $\chi^{\text{exp}}_{k_a}$ and $\chi^{\text{exp}}_{\text{base}}$. The solution $S^*(\omega)$ to Eq.~(\ref{CSconvec-0}) is an accurate approximation of (the discretized approximation of) $S(\omega)$, when the number of generated Fourier functions satisfies $m \geq {\Omega}(s\log N)$. Rigorous error bounds for $S^*(\omega)$ can be derived using standard techniques from compressed sensing theory \cite{candes2011probabilistic}.

In practice, one may have prior information about the range of frequencies $[f_1,f_2]$ where noise will occur. In the above discussion, we have considered the special case where $f_1 = 0$. However, one can easily extend our method to the case where $f_1 > 0$, in the following way: First, one constructs random pulse sequences whose window functions are sinusoidal functions with wavelengths $(f_2-f_1)/k$, for randomly chosen integers $k \in \lbrace 1,2,3,\ldots \rbrace$. Then, when solving for the noise spectrum $S(\omega)$, one imposes constraints that specify the values of $S(\omega)$ for $\omega \in [0,f_1]$. Compressed sensing then works in the same way described above. When the range $[f_1,f_2]$ is large, compressed sensing is expected to outperform conventional CPMG pulse sequences.

Fig.~\ref{fig:cs2} presents a numerical simulation of the CS method on an ideal sparse spectrum. The solid blue line represents a 2-sparse spectrum with $N=250$ grid points. The red circles represent the reconstructed spectrum, $S^*(\omega)$, obtained from CS with $m=12$ different Fourier basis functions.  By accurately identifying the non-zero elements of the original spectrum (blue line),  the CS method (red circles) succeeeds in reconstructing it.

We further examine the accuracy of CS in reconstructing ideal spectra under different sparsities, $s$,  as a function of the number of Fourier basis functions, $m$ (Fig.~\ref{fig:csphase}). For each sparsity (different curves in Fig.~\ref{fig:csphase}), we randomly generate 200 sparse spectra with $N=250$ grid points (see Appendix \ref{sec-random-sparse-spectra}) to obtain the averaged accuracy. The accuracy is defined as the $L_\infty$ norm of the difference between the discretized true spectrum, $S(\omega)$, and the reconstructed spectrum, $S^*(\omega)$. For each sparsity, the accuracy undergoes a clear phase transition at a certain value of $m$. We define $m_c$ as the critical number of Fourier basis functions for which  the accuracy  reaches 0.5 (e.g., for $s=13$, $m_c=40$). It can be seen that $m_c$ is a linear function of $s$ (inset of Fig.~\ref{fig:csphase}), which is consistent with the theoretically-predicted proportionality between $m$ and $s\log N$.

\subsection{Application to Quantum Dots}

Next, to quantify the performance of the CS method for realistic physical systems, we explore the ability of the method to extract the spectral density of noise that interacts with InAs/GaAs quantum dots. This noise represents the decoherence of the quantum dots due to their hyperfine interaction with an ensemble of nuclear spins broadened by strain \cite{stockill2016quantum, farfurnik2021all}. The solid blue line in Fig.~\ref{fig:FIG6a} shows the theoretical spectral density of such a  noise source calculated from the Fourier transform of the autocorrelators of the fluctuating nuclear spins, while considering quantum dots of pure indium and arsenic at a temperature of 4 K and under a magnetic field of B = 2 T applied perpendicular to the growth direction of the dots (Voigt geometry) \cite{stockill2016quantum}. This spectrum consists of several narrow peaks at spectral frequencies that correspond to different Larmor frequencies of the nuclei. 

The red dots in Fig.~\ref{fig:FIG6a} represent the discrete spectrum obtained by simulating the performance of CS with $m=40$ different Fourier basis functions. 
While the theoretical spectrum is not ideally sparse, we adopt suitable data analysis techniques (least absolute
shrinkage and selection  (LASSO) \cite{hastie2009elements}, along with cross-validation (CV), see Appendix \ref{sec-lasso}) to successfully identify the centers of the major peaks.

We quantify the accuracy of extracting the InAs/GaAs noise spectrum by comparing the central frequencies of the largest three peaks obtained from CS to their theoretical values. The theoretical central frequencies, $\hat{\omega}^{\text{th}}=(\omega_1, \omega_2, \omega_3)$, are calculated by a Gaussian fitting to the theoretical spectrum. The experimental results, $\hat{\omega}^*=(\omega_1^*, \omega_2^*, \omega_3^*)$, are the weighted mean values of the frequencies from the neighboring non-zero discrete $S^*(\omega)$ obtained from CS. The reconstruction accuracy is defined as the $\ell_\infty$ norm of the difference between these two vectors, i.e. $|\hat{\omega}^*-\hat{\omega}^{\text{th}}|_{\infty}$. 

The solid blue line and the dashed red line in Fig. \ref{cscompare} represent the simulated reconstruction accuracies of CS with $N=167$ and $N=667$ grid points, respectively.  In these simulations, we assume no experimental errors  and only focus on the effect of the number of different sets of experiments, $N_\text{set}$. For CS, $N_{\text{set}}=m+1$ for $m$ Fourier basis functions and one additional experiment with the base random pulse sequence. For both choices of $N$,  we observe a sharp change (phase transition) in the accuracy of reconstructing the spectrum at $N_{\text{set}} \approx 40$. The reconstruction accuracies then converge to constant values inversely proportional to $N$ (e.g., the dashed red line for $N=667$ has a lower baseline than the solid blue line for $N=167$). As a result, increasing the number of grid points $N$ 
in post-processing could boost the spectral resolution of CS without adding any resources.

To further demonstrate the resource efficiency of the CS method, we compare the accuracies of CS in resolving the InAs/GaAs noise spectrum to the ones obtained by using the conventional CPMG method (dotted green line in Fig.~\ref{cscompare}). For CPMG, the number of sets of experiments required for noise spectroscopy is given by $N_\text{set}={2\omega_c T}/{\pi}$, i.e., the number of different sequences that probe the noise spectrum over the frequency range $[0, \omega_c]$ given the total experiment time, $T$ \cite{haeberlen1976advances,bylander2011noise,wang2021intrinsic,bar2012suppression,malinowski2017notch,chan2018assessment,yuge2011measurement,farfurnik2021all}. 

For $N_\text{set} < 100$, the CPMG sequences cannot resolve adjacent spectral peaks of the InAs/GaAs noise spectrum (i.e., $|\hat{\omega}^*-\hat{\omega}^{\text{th}}|_{\infty} > 1$), because the sampling frequency interval associated with the sequences is wider than the spectral distance between nearby peaks in the spectrum. Meanwhile, for $N_\text{set} > 100$, the CPMG protocol can resolve the desired spectrum, with accuracy inversely proportional to $N_\text{set}$. However, to achieve a certain level of accuracy, the CPMG sequences require at least an order of magnitude more resources than the CS method. For example, as we illustrate for the InAs/GaAs noise spectrum, $|\hat{\omega}^*-\hat{\omega}^{\text{th}}|_{\infty} \approx 0.2$ for CS dashed red line) with $N=667$ grid points and $N_\text{set}\approx 40$; but this accuracy is hardly reached by CPMG (dotted green line) up to $N_\text{set}\approx 500$.



\subsection{Experimental Demonstration}

Finally, we demonstrate the compressed sensing technique on two experimental quantum devices. This provides evidence that our methods can be made robust to various imperfections that occur in real experiments. 

For example, in real experiments, the pulses are imperfect as they have finite pulse width and often contain some systematic error. Using our method, the effects of the finite pulse width will be felt equally across the frequency domain; whereas using CPMG pulse sequences, these effects will be worse at high frequencies and less harmful at low frequencies. It is not surprising that our method behaves this way, because our method works across a broad range of frequencies, i.e., it is a ``broadband'' measurement technique. 

As for systematic errors like over-rotation or axis error, it is well known in the context of CPMG pulse sequences that these kinds of errors can be canceled out by using composite pulses \cite{choi2020robust}. The same technique can be applied to our random pulse sequences. In the following experiments,  we apply robust $\pi\bar{\pi}$ pulse sequences (each $\pi$ pulse followed by a $-\pi$ pulse).

\begin{figure}
\begin{centering}
\subfloat[]{\label{fig:ibm}%
  \includegraphics[width=0.8\textwidth]{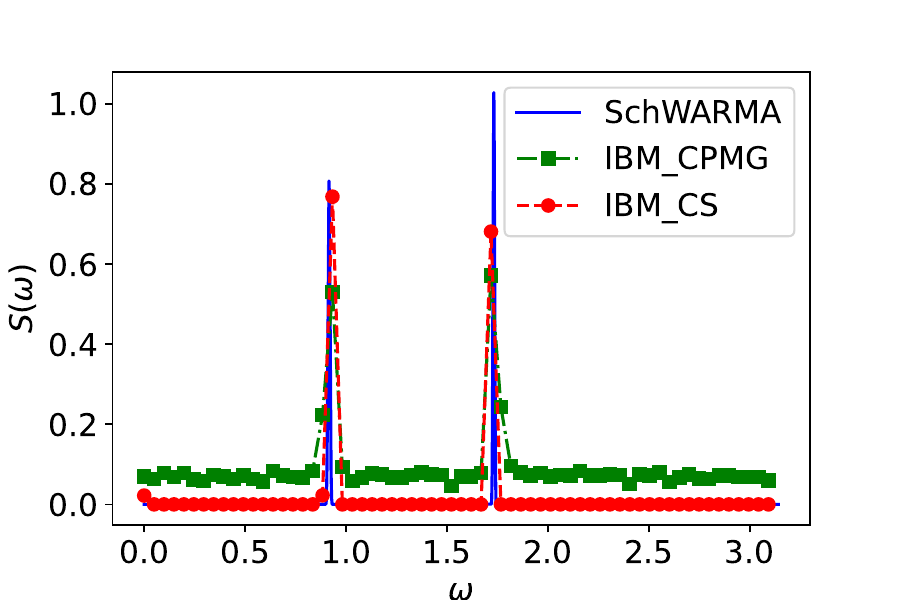}
}
\par\end{centering}
\begin{centering}
\subfloat[]{\label{fig:ibmtrans}%
  \includegraphics[width=0.8\textwidth]{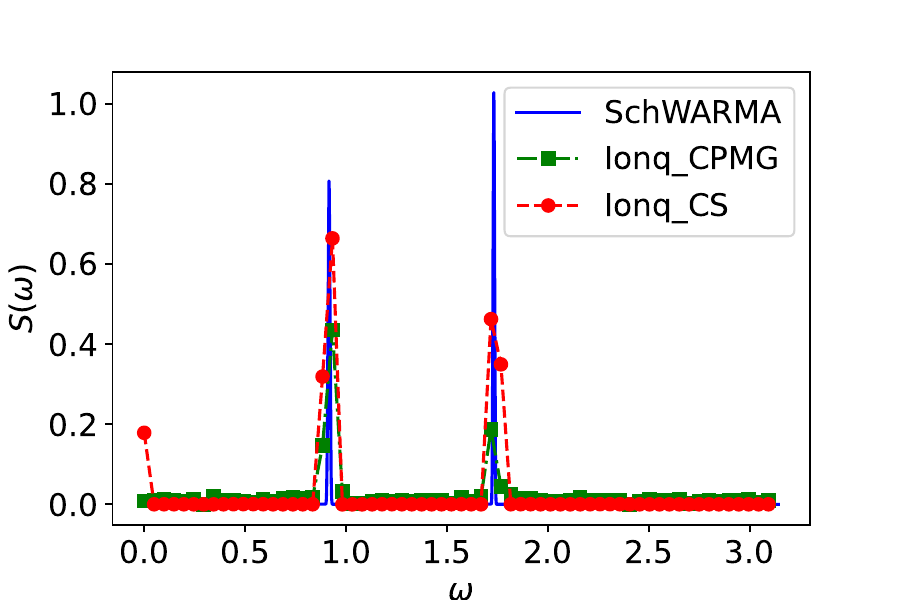}%
}
\par\end{centering}
\begin{centering}
\subfloat[]{\label{phtrans}%
  \includegraphics[width=0.8\textwidth]{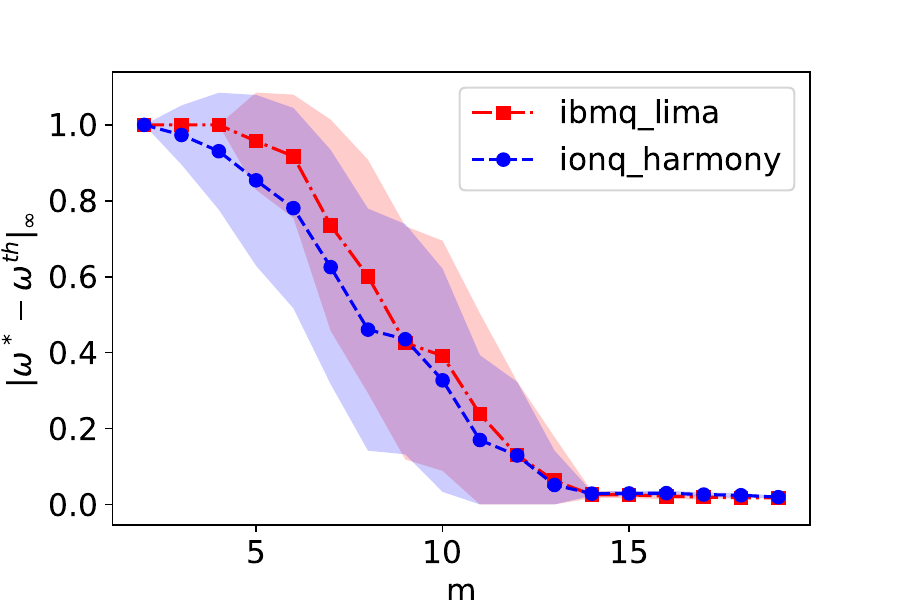}%
}
\par\end{centering}
\caption{(a) A reconstruction of the (artificially generated) sparse noise spectrum on the super-conducting qubit $ibmq\_lima$. The blue solid line represents the ideal sparse spectrum we want to inject using SchWARMA, with maximum intensity normalized to 1. The green squares represent the CPMG (plus NNLS) result with $64$ sets of experiments, each repeated $5000$ times ($(M,N_\text{CPMG})=(64,5000)$). The red circles represent the CS reconstruction with $m=10$ different Fourier basis functions and random pulse sequences with $(M, N_1, N_2)=(64, 100, 50)$. (b) Same experiments performed on $ionq\_harmony$. (c) The accuracy of reconstructing the central frequencies of the injected noise spectrum on different platforms as a function of the number of sets of experiments.  The dashed red and blue lines represent the results for $ibmq\_lima$ and $ionq\_harmony$, respectively. The shaded areas represent $\pm 1$ standard deviation, averaged over 60 different choices of CS bases.}
\end{figure}

The first device we examined is $ibmq\_lima$, a superconducting quantum information processor built by IBM. We utilize the Schrodinger Wave Autoregressive Moving Average (SchWARMA) tools \cite{schultz2021schwarma, murphy2022universal} to inject noise (with a sparse noise spectrum) into the qubit. The solid blue line in Fig.~\ref{fig:ibm} shows a random $2$-sparse spectrum. Noise with this spectrum was generated using SchWARMA. The green squares represent the spectrum reconstructed from the CPMG method, plus a non-negative least squares fitting (NNLS). The CPMG process contains $64$ sets of experiments on different frequencies, with each experiment repeated $5000$ times ($(M,N_\text{CPMG})=(64,5000)$). The red circles represent the reconstructed spectrum obtained from CS with $m=10$ different Fourier basis functions, each contains random pulse sequences with $(M, N_1, N_2)=(64, 100, 50)$. We also use LASSO and CV to analyze the data. 

Compared to the CPMG method, the CS result captures the main peaks of the injected noise, while eliminating the background white noise caused by the gate and SPAM (state preparation and measurement) errors. This is as expected, because the random pulse sequences we used here were designed to search for sparse ``pulse-like'' features that resemble delta functions, rather than smoothly-varying background noise such as white noise or 1/f noise \footnote{In principle, it is possible to design random pulse sequences with window functions that are more suitable for parametric fitting of 1/f noise. It is an interesting question whether this approach could perform better than parametric fitting of 1/f noise using CPMG pulse sequences.}.

\begin{table}[h]
    \caption{\label{tab:lima}Properties of qubit $0$ in $ibmq\_lima$ and average performance in $ionq\_harmony$ \cite{ionq}. SPAM represents error associated with state preparation and measurement inaccuracies. Gate error represents averaged error rate occurring during the operation of single-qubit gates.}
    \begin{ruledtabular}
        \begin{tabular}{cccccc}
            device & $T_1$ & $T_2$ & SPAM& Gate error& Gate time \\
            \hline
            $ibmq\_lima$ & 109 $\mu s$ & 188 $\mu s$
& 0.66\%
& 0.14\%
 &305.77 $ns$\\
\hline
            $ionq\_harmony$ &10 $s$ & 0.2 $s$
& 0.26\%
& 0.15\%
& 10 $\mu s$
\\

        \end{tabular}
    \end{ruledtabular}
\end{table}

We also perform the same experiments on $ionq\_harmony$, a trapped-ion quantum computing platform built by IonQ. The basic settings, including sets of experiments and numbers of pulse sequences, are exactly the same as the ones for $ibmq\_lima$. The results are shown in Fig.~\ref{fig:ibmtrans}. The CS method again well captured the main peaks of the injected noises. Compared with the superconducting qubits result, the trapped ion experiments have less background noise in the CPMG, due to the smaller SPAM error rate (see table \ref{tab:lima}). 

Similar to Fig.~\ref{cscompare}, we can further quantify the accuracy of the CS method on both platforms by comparing the central frequencies of the peaks obtained from CS to the actual values.  The dashed red and blue lines in Fig.~\ref{phtrans} represent the results for $ibmq\_lima$ and $ionq\_harmony$. It is interesting to see that the phase transition points of the accuracy are almost the same for both platforms, despite different hardware properties.

\section{Outlook}
To conclude, we develop a new method for qubit noise spectroscopy based on the realization of random pulse sequences. By using mathematical techniques for phase retrieval, this method allows us to measure arbitrary linear functionals of the noise spectrum. As an application, we demonstrate the reconstruction of sparse noise spectra, by using random pulse sequences together with compressed sensing (CS). Furthermore, the proposed method can be used to reconstruct noise spectra of realistic physical systems, such as optically-active quantum dots, with an order of magnitude less resources than conventional dynamical decoupling techniques. 

While it may seem surprising that random pulse sequences can be used to implement such a large class of spectroscopic measurements, our results are consistent with recent findings on the realization of optimal quantum control using switching pulse sequences \cite{chen2023quantum}. These results suggest that optimal control and characterization of large numbers of qubits can be achieved using relatively simple control electronics (i.e., without resorting to arbitrary waveform generators). 

For future research, our method can be generalized to incorporate pulses with durations other than $\pi$, and toward the characterization of multi-qubit systems \cite{paz-silva2017multiqubit}. 
In addition, while we have only considered the reconstruction of the noise spectrum on a finite set of sample points, similar ideas can be applied to reconstruction over continuous domains \cite{candes2014towards, 6576276, chi2020harnessing}.

Finally, the accuracy of CS utilizing random pulse sequences strongly depends on the spectral properties of the probed noise source. Beyond the experimental realization of CS on the specific platform of InAs/GaAs quantum dots, it would be beneficial to study the performance of random pulse sequences on realistic noise sources with various spectral features. For example, our method can be used to probe noise spectra of NV centers in diamond, which consist of a sharp (sparse) peak due to the interaction with 13C nuclear spin as well as slow decaying components associated with a bath of P1 electronic spins \cite{romach2019measuring}.

\section{Acknowledgements}

It is a pleasure to thank Edo Waks, Gregory Quiroz, Kevin Schultz, and others at the Johns Hopkins University Applied Physics Laboratory, for helpful discussions. 
In addition, we thank an anonymous referee for thoughtful comments on an earlier version of this paper. 
This work was partially supported by an AFOSR MURI on Scalable Certification of Quantum Computing Devices and Networks, a U.S. Department of Energy ASCR ARQC award on Fundamental Algorithmic Research for Quantum Computing (FAR-QC), and Office of Science, National Quantum Information Science Research Centers, Quantum Systems Accelerator, and the NSF QLCI for Robust Quantum Simulation (RQS), and NSF-IMOD STC Center DMR-2019444. A.S. acknowledges support from a Chicago Prize Postdoctoral Fellowship in Theoretical Quantum Science. Any mention of commercial products is for information only, and does not indicate endorsement by NIST.

\newpage
\bibliographystyle{ieeetr}
\bibliography{reference}

\appendix
\pagebreak
\widetext

\section{Noise spectroscopy for the spin-boson model}
\label{sec-spin-boson-model}

This section will demonstrate how to extend the noise spectroscopy method described in this paper to characterize \textit{quantum} environments. For example, in the spin-boson model \cite{viola1998dynamical, uhrig2007keeping, uhrig2008exact}, the system qubit is weakly coupled with many harmonic oscillators. The bath Hamiltonian and the coupling between system and bath can be written as
\begin{equation}
    \hat{H}_B + \hat{H}_V=\sum_k \omega_k a^\dagger_k a_k - \frac{1}{2}\sum_k (g_ka_k + g_k^*a^\dagger_k)\sigma_z,
\end{equation}
where $\hat{H}_B, \hat{H}_V$ are the bath and interacting Hamiltonian, $a_k (a^\dagger_k)$ is the $k$th mode annihilation (creation) operator of the bath, and $g_k$ is the coupling strength between $k$th mode and the qubit. The sum over $k$ can be approximated by the frequency integral
\begin{equation}
    \sum_k |g_k|^2 \approx \int_0^\infty J(\omega) d\omega,
\end{equation}
where $J(\omega)$ is the spectral function of the bath. When the bath is in thermal equilibrium at inverse temperature $\beta$, the noise spectrum can be calculated as
\beq
\so=\begin{cases}
    \pi J(\omega)(\coth(\beta\omega/2) +1), \omega > 0,\\
    \pi J(-\omega)(\coth(-\beta\omega/2) -1), \omega < 0.
\end{cases}
\eeq
Unlike a classical bath, the spectrum $S(\omega)$of the bosonic bath contains ``classical parts" and ``quantum parts"\cite{paz-silva2017multiqubit},
\beq
\begin{aligned}
    &S^+(\omega) = (\so+S(-\omega))/2,\\
    &S^-(\omega) = (\so-S(-\omega))/2.\\
\end{aligned}
\eeq
The ``quantum parts", $S^-(\omega)$, is nonzero only when the bath is quantum. However, we are only able to characterize the ``classical parts" using the spectroscopy method we discussed in this paper, as the decay rate only depends on $S^+(\omega)$, 
\beq
\chi(T) = \int_{-\infty}^{\infty}\frac{d\omega}{2\pi}S^+(\omega)W(\omega).
\eeq
This brings difficulty to fully characterize $S(\omega)$. For example, suppose we are interested in measuring a linear functional $I=\infint S(\omega) T(\omega) d\omega$, where the target function $T(\omega)$ is not an even function of $\omega$. We may as well define the even and odd parts of $T(\omega)$ in the following way,
\beq
\begin{aligned}
    &T^+(\omega) = (T(\omega)+T(-\omega))/2,\\
    &T^-(\omega) = (T(\omega)-T(-\omega))/2.\\
\end{aligned}
\eeq
The noise spectroscopy method only allows us to probe $I^+=\infint S^+(\omega) T^+(\omega) d\omega$, while the information about the quantum part, $I^-=\infint S^-(\omega) T^-(\omega) d\omega$, is missing. However, when the bath is at thermal equilibrium, this difficulty can be circumvented by observing that
\beq
\begin{aligned}
    &S^-(\omega) = S^+(\omega)  \tanh{(\beta\omega/2)}. 
\end{aligned}
\eeq
Thus, if we know the inverse temperature $\beta$, we can set up a new target function $T^*(\omega)$ that is always even on $\omega$,
\beq
T^*(\omega)=T^+(\omega)+T^-(\omega)\tanh{(\beta\omega/2)}. 
\eeq
Then, we can then use the random pulse method to measure the following linear functional
\beq
\begin{aligned}
I^*&=\infint S(\omega) T^*(\omega) d\omega\\
&= \infint S^+(\omega)  T^+(\omega) d\omega+ \infint S^+(\omega)  \text{tanh}(\beta\omega/2) \cdot T^-(\omega)d\omega\\
&=\infint S^+(\omega)  T^+(\omega)d\omega + \infint S^-(\omega) T^-(\omega)d\omega\\
&=\infint \so T(\omega) d\omega=I.
\end{aligned}
\eeq
That is to say,  we can fully characterize the spectrum of a quantum spin-boson model in thermal equilibrium, provided we know the temperature of the bath.

\section{Dynamical decoupling for noise spectroscopy}
\label{sec-cpmg-2nd-order-corrections}
Here we introduce the deconvolution procedure we used to reconstruct the noise spectrum\cite{bar2012suppression, romach2019measuring, farfurnik2021all}. The window function of the CPMG sequence equals
\beq
\label{dd0}
\begin{aligned}
W(\omega)=\begin{cases} \frac{32}{\omega^2}\sin^4(\tfrac{\omega \tau }{4})\sin^2(\tfrac{\omega M\tau }{2})/\cos^2(\tfrac{\omega \tau }{2}), \text{for even $M$}\\
\frac{32}{\omega^2}\sin^4(\tfrac{\omega \tau }{4})\cos^2(\tfrac{\omega M\tau }{2})/\cos^2(\tfrac{\omega \tau }{2}), \text{for odd $M$}.
\end{cases}
\end{aligned}
\eeq
This window function contains one major peak, at $\omega_0 \approx \tfrac{\pi}{2}\tau$, and multiple minor peaks at higher harmonics. The reconstructed spectrum
$S(\omega)$ from the decoherence data is a solution to the Fredholm type equation
\begin{equation}
    \begin{aligned}
        \chi(T) = \int_{-\infty}^{\infty} \frac{d\omega}{2\pi}S(\omega)W(\omega).
    \end{aligned}
\end{equation}
Since we assume that the spectrum decays to zero at high frequency, we can reconstruct the spectrum from higher to lower frequencies by subtracting
the effect of higher harmonics recursively using the analytical form of $W(\omega)$ in Eq. (\ref{dd0}).

The detailed algorithm goes in the following way: We first approximate the spectrum from the main contribution of the decay exponent due to the major peak, 
\begin{equation}
    \begin{aligned}
        \chi(T) & \approx \int_{\omega_-}^{\omega_+} \frac{d\omega}{2\pi}S(\omega)W(\omega)\\
        & \approx \frac{S(\omega_0)}{2\pi}\int_{\omega_-}^{\omega_+}d\omega W(\omega)\\
        & \approx \frac{S(\omega_0)}{2\pi} \Sigma,
    \end{aligned}
\end{equation}
where $\Sigma$ is the area under the major peak, defined as the integral of $W(\omega)$ between the nearest two local minima $\omega_-$ and $\omega_+$ around $\omega_0$. Thus we get the first-order approximation of the spectrum,
\begin{equation}
\label{ddfirst}
    S_1(\omega_0) \approx \frac{2\pi\chi(T)}{\Sigma}.
\end{equation}

We then apply different CPMG sequences to sample the spectrum at various frequencies. We sort all sampled frequencies from highest to lowest, and label them as $[\omega_0^1,\omega_0^2, \ldots, \omega_0^\text{max}]$. The corresponding window functions are labelled as $[W^1(\omega), W^2(\omega), \ldots, W^\text{max}(\omega)]$. Assuming that the noise spectrum vanishes above the highest sampled frequency, $\omega_\text{max}$, expression (\ref{ddfirst}) is exact such that $S(\omega^\text{max}_0)=S_1(\omega^\text{max}_0)$. Next we can iteratively calculate the spectrum from the highest frequency downwards,
\begin{equation}
    S(\omega^i_0)=S_1(\omega_0^i)-\int_{\omega^i_+}^{\infty}\frac{d\omega}{2\pi}S^*(\omega)W^i(\omega),
\end{equation}
where $S^*(\omega)$ takes the value of the previously (recursively) extracted spectrum at the sampled frequency closest to $\omega$.

Another commonly applied reconstruction method for noise spectroscopy is the Alvarez-Suter method \cite{ alvarez2011measuring}, which requires the assumption of an infinite number of pulses to justify the comb approximation. Compared to that, the decomposition method mentioned above utilize the exact form of the window function, and in general works better for the spectra decaying to zero at high frequency \cite{bar2012suppression, szankowski2018accuracy, szankowski2019transition}.


\section{Methods for generating random pulse sequences}
\label{sec-generating-random-pulse-sequences}

In this section, we will introduce two methods for constructing suitable finite impulse response (FIR) filters, in order to produce random pulse sequences whose window functions will approximate a desired target function, as described in Section \ref{sec-random-pulse-sequences}. The first method is very general, and is capable of approximating any desired target function. The second method is more specialized, and is capable of approximating sinusoidal target functions, which are useful for compressed sensing, as described in Section \ref{sec-compressed-sensing}.

\subsection{General method}
\label{sec-generating-random-pulse-sequences-1}
We first generate a sequence of $M$ independent Gaussian random variables, denoted $\vec{N}  = (N_0, N_1, \ldots, N_m, \ldots, N_{M-1})$. Next, we apply a finite impulse response (FIR) filter with suitably chosen coefficients $(a_0, a_1, \ldots, a_{\lambda-1})$, and then apply the sign function, to get the random variables representing the random pulse sequence, $\vec{U} = (U_0, U_1, \ldots, U_{M-\lambda})$. In other words,   
\begin{equation}
U_m=\text{sign}(\sum_{i=0}^{\lambda-1} a_i N_{m+i}), \quad m=0,1,2,\ldots,M-\lambda. 
\end{equation}

\begin{figure}[h!]
\includegraphics[width=6cm]{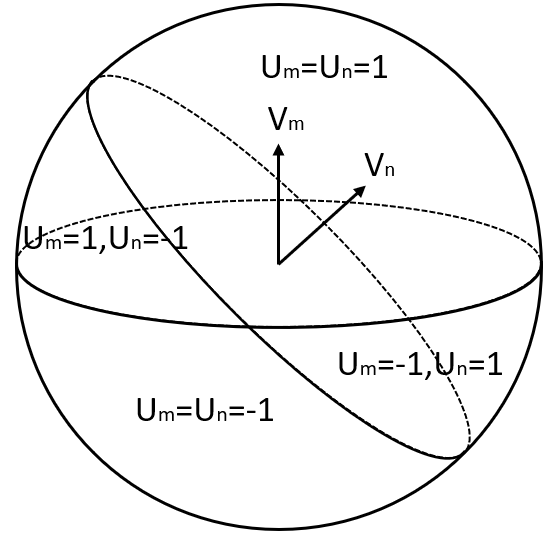}
\centering
\caption{A geometric interpretation of how to generate the random pulse sequences, $\vec{U}$, by hyper-plane tessellation. The space is separated by hyper-planes with normal vectors indicated by the FIR coefficients, $\vec{V}_m$. Every time we generate a Gaussian random sequence $\vec{N}$, we can calculate $\vec{U}$ by checking which region $\vec{N}$ lies in after the tessellation. For example, if $\vec{N}$ lies in the right upper corner of the hyper sphere we show, then $U_m=U_n=1$. Since the elements in $\vec{N}$ are independent Gaussian random variables, $\vec{N}$ is uniformly distributed over all directions. Thus, the covariance between different $U_m$ can be calculated by the solid angles between these hyper-planes.}
\label{fig:geo}
\end{figure}

Without loss of generality, we can extend the FIR coefficients to a normalized $M$ dimensional vector $\vec{V}_m=(0, 0, \ldots, a_0, a_1, \ldots, a_{\lambda-1}, 0, \ldots, 0)$ where $a_0$ appears in position $m$, so that $U_m  = \text{sign}(\vec{V}_m\vec{N}^T)$ and $\vec{V}_m\vec{V}_m^T=1$. Furthermore, the whole construction process can also be viewed geometrically as a hyper-plane tessellation, see Fig.~\ref{fig:geo}. $\vec{V}_m$ is the normal vector of a hyper-plane in $\mathbb{R}^{M}$ which divides the entire space into two regions. The value of $U_m$ is determined by whether the random vector $\vec{N}$ lies within the corresponding region. The covariance $R(k)$ between $U_m$ and $U_{m+k}$ (for $k\geq 1$) can thus be calculated using the angle between different hyper-planes:
\beq
\begin{aligned}
\label{rkderive}
R(k) & = \EE( U_{m+k}U_m)\\
&= P(U_m\neq U_{m+k})\cdot(-1)+P(U_m = U_{m+k})\cdot(+1)\\
& =\frac{\arccos{(\frac{\vec{V}_{m}\vec{V}_{m+k}^T}{|\vec{V}_{m}||\vec{V}_{m+k}|})}}{\pi}\cdot(-1)+\frac{(\pi-\arccos{(\frac{\vec{V}_{m}\vec{V}_{m+k}^T}{|\vec{V}_{n}||\vec{V}_{m+k}|})}}{\pi}\cdot(+1)\\
      & =\frac{2}{\pi}\arcsin{({\vec{V}_{n}\vec{V}_{n+k}^T})}=\frac{2}{\pi}\arcsin{(\sum_{i=0}^{\lambda-1-k} a_i a_{i+k})}.
\end{aligned}
\eeq

Next, we can calculate the expectation value of the window function generated by the random pulse sequences.
Recall that the decay exponent equals
\beq 
\label{equq}
\chi_U(T) = \frac{1}{2\pi} \int_{\pi/\tau}^{\pi/\tau} d\omega S(\omega) W(\omega),
\end{equation}
with
\begin{equation}
W_{U} (\omega)= \abs{\tilde{f}_{U}(\omega)}^2,
\end{equation}
\begin{equation}
\tilde{f}_{U}(\omega) = \int_{-\infty}^\infty dt f_U(t) e^{i\omega t},
\end{equation}
\begin{equation}
f_U(t) = \begin{cases}
U_m &\text{ if $t \in [(m-1)\tau, m\tau)$, $m \in \set{1,2,\ldots,M}$,} \\
0   &\text{ otherwise.}
\end{cases}
\end{equation}
Hence we have 
\begin{equation}
\begin{split}
W(\omega) &= \abs{\tilde{f}_U(\omega)}^2 \\
&= \tau^2 \sinc^2(\tfrac{\omega\tau}{2}) \Bigl| \sum_{m=1}^M U_m e^{i\omega m\tau} \Bigr|^2 \\
&= \tau^2 \sinc^2(\tfrac{\omega\tau}{2}) \vec{U}^T A(\omega) \vec{U}, 
\end{split}
\end{equation}
where we define the matrix $A(\omega) \in \CC^{M\times M}$ whose $(m,m')$ entry is 
\begin{equation}
A_{m,m'}(\omega) = e^{i\omega (m-m')\tau}.
\end{equation}

We can further define the matrix $B(S) \in \CC^{M\times M}$ whose $(m,m')$ entry is 
\begin{equation}\label{bdef}
B_{m,m'}(S) = \hat{S}_{m-m'} \equiv \int_{-\pi/\tau}^{\pi/\tau} d\omega S(\omega)\sinc^2(\tfrac{\omega\tau}{2}) A_{m,m'}(\omega)
\end{equation}
where $\hat{S}_{m-m'}$ is equal to the $(m-m')$'th coefficient of the Fourier series expansion of $S(\omega)\sinc^2(\tfrac{\omega\tau}{2})$.
Thus we can write $\chi_U$ in the following form:
\begin{equation}\label{eqn-QT}
\chi_U = \frac{\tau^2}{2\pi} \vec{U}^T B(S) \vec{U}, 
\end{equation}

Therefore, the expectation value of $W(\omega)$ is 
\beq 
\label{wfunc}
\begin{aligned}
\EE(W(\omega)) &=\tau^2 \sinc^2(\tfrac{\omega\tau}{2})\EE(\vec{U}^T A(\omega) \vec{U})\\
&= \tau^2 \sinc^2(\tfrac{\omega\tau}{2})\sum_{m,m'}(\EE(U_mU_{m'})A_{m,m'}(\omega))\\
&=M\tau^2\sinc^2(\tfrac{\omega\tau}{2})[1+2\sum_{k=1}^{\lambda}R(k)\cos{(k\omega\tau)}(1-\tfrac{k}{M})].
\end{aligned}
\eeq 
The expectation value of $\chi_U(T)$ is
\beq
\EE(\chi_U(T))=\frac{M\tau^2}{2\pi}\int_{-\omega_c}^{\omega_c} d\omega S(\omega) \sinc^2(\tfrac{\omega\tau}{2})[1+2\sum_{k=1}^{\lambda}R(k)\cos{(k\omega\tau)}(1-\tfrac{k}{M})]
\eeq

\subsection{Approximating a desired window function}
\label{sec-generating-random-pulse-sequences-2}

Recall from Section \ref{sec-random-pulse-sequences} that our goal is to ensure that the averaged window function $\EE(W(\omega))$ approximates some prescribed target function $T(\omega)$. Note that the cosine functions $\{\cos(k\omega\tau)\}$ form an almost complete basis (the zeroth term excluded) in the region $[-\tfrac{\pi}{\tau}, \tfrac{\pi}{\tau}]$. Thus we take the following approach: we match the time interval between segments and the cutoff frequency of the noise (i.e., setting $\tau=\tfrac{\pi}{\omega_c}$), and we adjust the random pulse generator (i.e., optimizing the filter coefficients $(a_0, a_1, \ldots, a_{\lambda-1})$), so that 
\beq 
\label{rks}
R(k) =\tfrac{M}{\pi(M-k)} \int_{-\omega_c}^{\omega_c}\frac{cT(\omega)\cos(k\omega\tau)}{\sinc^2(\tfrac{\omega\tau}{2})}d\omega\quad (\forall k=1,\ldots,\lambda),
\eeq 
which implies
\beq 
\label{ews}
\EE(W(\omega)) \xrightarrow{\lambda\rightarrow \infty} M\tau^2[cT(\omega)+(1-cT_0)\sinc^2(\tfrac{\omega\tau}{2})].
\eeq 
Here $c$ is an adjustable parameter that ensures the positivity  of $W(\omega)$, and $T_0$ is a constant term depending on $T(\omega)$,
\beq 
\label{t0s}
T_0 = \tfrac{1}{\omega_c} \int_{-\omega_c}^{\omega_c}\frac{T(\omega)}{\sinc^2(\tfrac{\omega\tau}{2})}d\omega.
\eeq 

Combining equations (\ref{rkderive}) and (\ref{rks}), we finally get an equation to generate the filter coefficients:
\beq
\label{1dphr}
\sum_{i=0}^{\lambda-1-k} a_i a_{i+k} = \sin{(\frac{\pi}{2}R(k))}=\sin{(\tfrac{ M}{2(M-k)} \int_{-\omega_c}^{\omega_c}\frac{cT(\omega)\cos(k\omega\tau)}{\sinc^2(\tfrac{\omega\tau}{2})}d\omega}) \quad (\forall k=1,\ldots,\lambda).
\eeq
Note that the left side of this equation can be seen as the autocorrelation of a sequence formed by $a_i$, so solving for the values of $a_i$ is equivalent to a 1-dimensional discrete phase retrieval problem \cite{jaganathan2016phase}. This can be solved by switching to the Fourier domain, as follows.

Define $R(k)$ for all integers $k$, by defining $R(0) = 1$ and $R(k) = R(-k)$.
We define $q(m)$ to be a periodic function on the real line, with $q(m) = q(m+2\lambda-1)$, whose Fourier coefficients are $\sin{(\tfrac{ M}{2(M-k)} \int_{-\omega_c}^{\omega_c}\frac{cT(\omega)\cos(k\omega\tau)}{\sinc^2(\tfrac{\omega\tau}{2})}d\omega})$:
\beq\label{eqn-qm}
\begin{aligned}
q(m) & = \sum_{k=-(\lambda-1)}^{\lambda-1} \sin{(\tfrac{ M}{2(M-k)} \int_{-\omega_c}^{\omega_c}\frac{cT(\omega)\cos(k\omega\tau)}{\sinc^2(\tfrac{\omega\tau}{2})}d\omega}) e^{-k\frac{2\pi im}{2\lambda-1}}.
\end{aligned}
\eeq
Next, we define $b(m)$ to be a periodic function on the real line, with $b(m) = b(m+2\lambda-1)$, as follows:
\beq\label{eqn-bm}
\begin{aligned}
b(m) & =\sum_{j=0}^{\lambda-1} a_j e^{-j\frac{2\pi im}{2\lambda-1}}
\end{aligned}
\eeq
Then equation (\ref{1dphr}) becomes:
\begin{equation}
\label{ftphr}
q(m) = |b(m)|^2, \quad \forall m\in\mathbb{R}.    
\end{equation}
That is, given a function $q(m)$ (which is defined by (\ref{eqn-qm})), we want to solve for a function $b(m)$ that has the form (\ref{eqn-bm}).
Note that $m$ can take arbitrary real values. 

When $q(m)$ is non-negative, the existence of solutions to Eq.~(\ref{ftphr}) can be guaranteed by the well-known Fej\'er-Riesz theorem \cite{fejer-riesz, fejer1916trigonometrische}:
\begin{theorem}
\label{fejer}
(Fej\'er-Riesz) If 
\beq
Y(e^{it})=\sum_{n=-M}^{M} y(n)e^{-int}
\eeq
and assumes non-negative real values for all real t, then there is a polynomial
\beq
X(z)=\sum_{n=0}^{M}x(n)z^{-n}
\eeq
such that
\beq
Y(e^{it})=|X(e^{it})|^2.
\eeq
\end{theorem}

To use the Fej\'er-Riesz theorem, we need to guarantee the non-negativity of $q(m)$ for all $m$. We will do this by adjusting the value of $c$. 
For instance, we can always set $c$ sufficiently small so that $\max_{k \neq 0,\, |k|\leq\lambda-1}{|\sin{(\frac{\pi}{2}R(k))}|} \leq \frac{1}{2\lambda-2}$. Then we have
\beq
\begin{aligned}
q(m) & = 1+ \sum_{k\neq 0,\, |k|\leq\lambda-1} \sin{(\frac{\pi}{2}R(k))} e^{-k\frac{2\pi im}{2\lambda-1}}\\
& \geq 1+ \sum_{k\neq 0,\, |k|\leq\lambda-1} -|\sin{(\frac{\pi}{2}R(k))}| \geq 0,
\end{aligned}
\eeq
as desired.
This proves the existence of a solution to Eq.~(\ref{ftphr}), and thus Eq.~(\ref{1dphr}), for any target function $T(\omega)$. 

In practice, we would like to set $c$ as large as possible, and hence set $\max{|R(k)|}$ as large as possible (not necessarily bounded by $\max_{k \neq 0}{|\sin{(\frac{\pi}{2}R(k))}|}=\frac{1}{2\lambda-2}$).  That is to say, we want to find:
\beq
\label{cvalue}
c_\text{opt} = \text{arg max}  \:   c,   \text{ subject to}\:   q(m) \geq 0,\,  \forall m \in \mathbb{R}.
\eeq
An efficient and practical way to solve this problem is to relax these constraints, by restricting $m$ to lie in the set $\set{0,1,...,2\lambda-2}$, so that $e^{-k\frac{2\pi im}{2\lambda-1}}$ is uniformly distributed on the unit circle. Thus we can easily maximize $c$ subject to these constraints:
\beq
\label{crvalue}
c_\text{relax} = \text{arg max}  \:   c,   \text{ subject to}\:   q(m) \geq 0,\,  \forall m \in \{0,1,...,2\lambda-2\}.
\eeq
Since $c_\text{opt}$ is always smaller than $c_\text{relax}$, we should set $c$ to be slightly smaller than $c_\text{relax}$. In practice, this can be done by trial and error. 

Finally, we can solve Eq.~(\ref{ftphr}) to recover $b(m)$, and thus obtain the $a_i$. In practice, this can be done by solving the following nonlinear least-squares problem:
\beq
\begin{aligned}
\vec{a} = \text{arg min} \sum_{m=0}^{2\lambda-2} (q(m)-|b(m)|^2)^2,\, \text{such that } b(m)  =\sum_{j=0}^{\lambda-1} a_j e^{-j\frac{2\pi im}{2\lambda-1}},\, \forall m \in \{0,1,...,2\lambda-2\}.
\end{aligned}
\eeq
This can be solved using gradient descent methods, or the Gerchberg-Saxton (GS) algorithm \cite{jaganathan2016phase}.

\subsection{Alternative generator}
\label{sec-alternative-generator}
The general method is powerful since it can be used to approximate arbitrary target functions, $T(\omega)$,  using the average window function of the pulse sequences. However, in some special cases where the form of $T(\omega)$ is restricted, we can utilize other simplified but efficient pulse generators. For example, in the compressed sensing case, we want to generate random pulse sequences where only one of the covariances $R(k)$ ($k\neq 0$) is nonzero, i.e., there exists some $k'\geq 1$ such that, for all $k\geq 1$, $R(k)$ has the form $R(k) = \delta_{kk'}R(k')$. An alternative generator can be used for this particular job with a simpler structure and fewer hardware memories. 

Using the terminology of the previous section, this alternative method is equivalent to using a \textit{time-varying} FIR filter. The resulting covariances $\EE(U_i U_{i+k})$ depend on both the time $i$ and the time-difference $k$; but if one averages over $i$, one obtains the desired covariance $R(k)$. This kind of averaging occurs naturally in experiments, whenever the bath dynamics are \textit{time-independent}. 

\begin{figure}
\includegraphics[width=14cm]{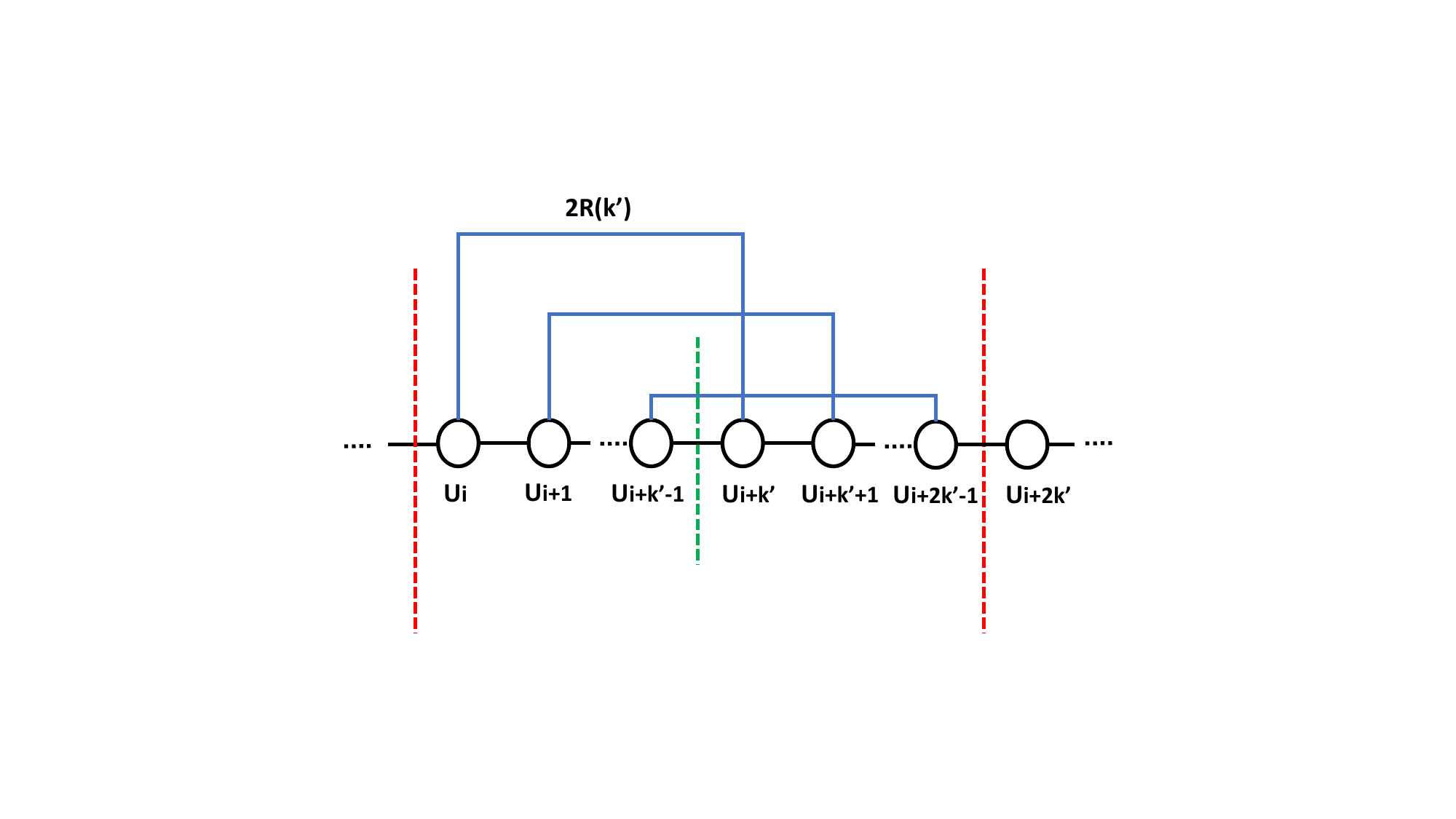}
\centering
\caption{An illustration to generate the random pulses with only one nonzero $R(k)$ (corresponding to $k=k'$) using the alternative generator. The black circles represent the random pulse sequence, labelled with random variables $\vec{U}$. The red dashed lines split the pulse sequence to individual segments with length $2k'$. The green dashed line further divides the segments into two halves (``left'' and ``right''). The random variables on the left side are correlated with corresponding ones on the right side, indicated by the blue lines. The distance between the correlated pairs is always $k'$, and the covariance is $2R(k')$. }
\label{generator2}
\end{figure}

The idea of this alternative method is to create correlated pairs. As shown in Fig. \ref{generator2}, the chain of circles represents the random pulse sequence, labelled by the random variables $\vec{U}=(U_1,U_2,...,U_M)\in\{1,-1\}^M$. The sequence of random variables is split into independent segments with length $2k'$, indicated by the red dashed lines, so that the random variables from different segments are independent. Each segment is further divided into two halves (``left'' and ``right'') by the green dashed lines. The variables on the left side of the segments are chosen uniformly random from $\{-1,1\}$. We then choose some $r$ in the range $[-1/2, 1/2]$. For every variable on the right side, one can always find a corresponding variable in the left side, such that the distance between them is $k'$. We denote the left and right variables in the pair as $U_i$ and $U_{i+k'}$. Conditioned on $U_i$, $U_{i+k'}$ is generated according to the following probability distribution:
\beq
\begin{aligned}
    &P(U_{i+k'}=u|U_i=u)=1/2+r,\\
    &P(U_{i+k'}=-u|U_i=u)=1/2-r,
\end{aligned}
\eeq
for all $u \in \set{1,-1}$.
As a result, the covariance between $U_i$ and $U_{i+k'}$ is $2r$. 

We assume that $M$ is divisible by $2k'$, and the pulse sequence starts with a complete segment. When one averages over all choices of $i$ (including those values of $i$ where $i+k'$ no longer lies in the same segment as $i$), the averaged covariance between $U_i$ and $U_{i+k'}$ is
\beq
\begin{aligned}
    R(k)&=\delta_{kk'}\EE_i(U_i U_{i+k'})\\
    &=\delta_{kk'}\frac{2r\cdot M/2}{M-k'}\\
    &=\delta_{kk'}\frac{M}{M-k'}r.
\end{aligned}
\eeq

Assuming $k' \ll M$, and substituting the result into Eq.~(\ref{wfunc}), we have
\beq
\EE(W(\omega)) =M\tau^2\sinc^2(\tfrac{\omega\tau}{2})[1+2r\cos{(k'\omega\tau)}].
\eeq
In the compressed sensing scenario, we can simply set $r=1/2$, which means that the paired variables take the same value. The expectation value of the window function is thus
\beq
\EE(W(\omega)) =M\tau^2\sinc^2(\tfrac{\omega\tau}{2})[1+\cos{(k'\omega\tau)}],
\eeq
which is perfect for compressed sensing (see Section \ref{sec-compressed-sensing}). Compared to the general method, this alternative method has a much simpler structure. In addition, those independent segments can be generated in parallel. This may be advantageous in the NISQ era where the hybrid quantum systems have limited memory resources.

\section{Accuracy of the random pulses method}
\label{sec-accuracy-details}
In this section, we will analyze the accuracy of the random pulse sequence method. Recall that we need to estimate the desired linear functional, $I=\int S(\omega)T(\omega)d\omega$, from the expected decay exponent $\EE(\chi)$ (Eq.~(\ref{chirps})). The accuracy of the estimator thus depends on the error in the decay exponent measured, $\chi^{\text{exp}}$.

\begin{theorem}
For random pulse sequence experiments containing $N_1$ different sequences with each repeated $N_2$ times, the accuracy of the measured decay exponent, $\chi^{\text{exp}} = \chi^{\text{exp}}_U(T)$, is upper bounded by
\begin{equation}
\begin{aligned}
    |\chi^{\text{exp}}_U(T)-\EE(\chi_U(T))| 
    = O\bigg(\frac{\sqrt{\text{Var}(\chi_U(T))}}{\sqrt{N_1}}
    + \frac{\mathbb{E}(\chi_U(T))}{\sqrt{N_1N_2}}\bigg),
\end{aligned}
\end{equation}
\begin{equation}\label{eqn-variance-upper-bound}
\begin{aligned}
    \text{Var}(\chi_U(T))\leq \frac{(7\tilde{\lambda}^2 + 10\tilde{\lambda} + 1)M\tau^3}{\pi}||S(\omega)\sinc^2(\tfrac{\omega\tau}{2})-
    \langle S(\omega)\sinc^2(\tfrac{\omega\tau}{2})\rangle||^2_{L2}.
\end{aligned}
\end{equation}
Here $\tilde{\lambda}$ is a parameter that quantifies the correlations in the random pulse sequence, which is defined precisely in Section \ref{sec-variance-decay-exponent-general-case}. $S(\omega)$ is the noise spectrum, $M$ is the total number of pulses, $\tau$ is the time between pulses, and $\langle \cdot \rangle$ denotes averaging over the frequency domain, i.e., $\langle f(\omega) \rangle = \frac{\tau}{2\pi} \int_{-\pi/\tau}^{\pi/\tau} f(\omega) d\omega$.
\end{theorem}

The proof of this theorem goes in this way: we first upper-bound the variance of the decay exponent, $\text{Var}(\chi_U(T))$, for a given group of random pulse sequences. Then we describe the method for estimating $\chi_U(T)$ (i.e., computing $\chi^{\text{exp}}_U(T)$) based on experimental data, and we derive a bound on the error in $\chi^{\text{exp}}_U(T)$.

\subsection{Variance of the decay exponent: the uncorrelated case}
\label{sec-variance-decay-exponent}
In this part we calculate the variance of the decay exponent. Using Eqs.~(\ref{bdef}) and  (\ref{eqn-QT}), the variance of $\chi_U(T)$  is given by
\beq
\label{varchi}
\begin{aligned}
\text{Var}(\chi_U(T)) & =\EE(\chi_U(T)^2)-\EE(\chi_U(T))^2\\
& = \frac{\tau^4}{4\pi^2}\sum_{ijkl}(\EE(U_iU_j U_k U_l)-\EE(U_i U_j)\EE(U_kU_l))B_{i,j}(S)B_{k,l}(S)\\
\end{aligned}
\eeq
The fourth order term $\EE(U_iU_jU_kU_l)$ is in general hard to evaluate. So let us first consider a special case, where the $U_i$ are completely independent. For this case, we will obtain the upper bound (\ref{eqn-variance-upper-bound}) with $\tilde{\lambda} = 0$. \\ 

This uncorrelated case is also referred in the main paper as  the base random pulse sequence. Recall that the random variables of the base sequence are all independent with $\EE(U_iU_j)=\delta_{ij}$, such that
\beq
\EE(U_iU_jU_kU_l)=\delta_{ij,kl}+\delta_{ik,jl}+\delta_{il,kj}-2\delta_{i,j,k,l}.
\eeq
Substitute it into Eq.~(\ref{varchi}). Combined with the fact that $B(S)=B(S)^T$ ($S(\omega)$ is an even function of $\omega$), we get
\begin{equation}
    \label{varzero}
    \begin{aligned}
        \text{Var}(\chi_U(T)) & =\frac{\tau^4}{4\pi^2} [2 \text{Tr}(B(S)^2)-2 \sum_i B_{i,i}(S)^2].\\
    \end{aligned}
\end{equation}

We can bound $\text{Tr}(B(S)^2)$ as follows, using Parseval's theorem:
\begin{equation}
    \begin{aligned}
        \text{Tr}(B(S)^2)&=\sum_{m,m'=1}^{M}|B_{m,m'}(S)|^2\\
        &=\sum_{m=1}^{M}\sum_{l=1-m}^{M-m}|\hat{S}_{l}|^2\\
        &\leq M\sum_{l=-\infty}^{\infty}|\hat{S}_{l}|^2\\
        &=\frac{2M\pi}{\tau}\int_{-\pi/\tau}^{\pi/\tau}S^2(\omega)\sinc^4(\tfrac{\omega\tau}{2})d\omega\\
        &=\frac{2M\pi}{\tau}||S(\omega)\sinc^2(\tfrac{\omega\tau}{2})||_{L2}^2.
    \end{aligned}
\end{equation}

We then have an upper bound for the variance of the decay exponent:
\begin{equation}
    \label{chibound}
    \begin{aligned}
        \text{Var}(\chi_U(T)) &=\frac{\tau^4}{4\pi^2}(\frac{4M\pi}{\tau}||S(\omega)\sinc^2(\tfrac{\omega\tau}{2})||_{L2}^2-2M||S(\omega)\sinc^2(\tfrac{\omega\tau}{2})||_{L1}^2)\\
        &\leq \frac{M\tau^3}{\pi}||S(\omega)\sinc^2(\tfrac{\omega\tau}{2})-\langle S(\omega)\sinc^2(\tfrac{\omega\tau}{2})\rangle||^2_{L2}.
    \end{aligned}
\end{equation}
Interestingly, the variance of the measured decay rate is bounded by the difference between the square of L2 norm and L1 norm of the function $S(\omega)\sinc^2(\tfrac{\omega\tau}{2})$. This can be seemed as the deviation of  $S(\omega)\sinc^2(\tfrac{\omega\tau}{2})$ from a constant function. When this function is exactly a constant function in the frequency regime $[-\pi/\tau,\pi/\tau]$, $\Var(\chi_U(T))$ is precisely 0. At the opposite extreme, if $S(\omega)\sinc^2(\tfrac{\omega\tau}{2})$ is a sparse signal with a single peak of width $\Delta \omega$, then $\Var(\chi_U(T))$ is of order $O(\tfrac{\omega_c}{M\Delta\omega}\EE^2(\chi_U(T)))$.


\subsection{Variance of the decay exponent: the general case}
\label{sec-variance-decay-exponent-general-case}

In this section, we will derive an upper bound on the variance of the decay exponent $\chi_U(T)$, in a more general case where $\EE(U_iU_j) = R(j-i)$ for some arbitrary correlation coefficients $R(1), R(2), R(3), \ldots$. 
For this purpose, we will define a parameter $\tilde{\lambda}$ that characterizes the structure of the correlations among the random variables $U_i (i = 1, \ldots , M)$. 

We will view $U_i$ as a Markov random field. According to Section \ref{sec-generating-random-pulse-sequences}, $U_i$ is defined by the sign of jointly normal random variables: $U_i=\text{sign}(\vec{V}_i \vec{N}^T)$. We define a dependency graph $G$ (on vertices $V=\{1, \ldots, M\}$) based on the normal random variables $\vec{V}_i \vec{N}^T$, in which vertices $i, j$ are connected if $\EE(\vec{V}_i \vec{N}^T \vec{V}_j \vec{N}^T) \neq 0$. It follows that $\vec{V}_i \vec{N}^T$ form a Gaussian Markov random field (GMRF) respect to $G$ \cite{speed1986gaussian}:
\begin{enumerate}
    \item For any $i, j$ not equal or adjacent, $\ConditionallyIndependent{\vec{V}_i \vec{N}^T}{\vec{V}_j \vec{N}^T}{\vec{V}_{V\backslash \{i, j\}} \vec{N}^T}$,
    \item For any $i\in V$ and $J \subset V$ not containing or adjacent to $i$, $\ConditionallyIndependent{\vec{V}_i \vec{N}^T}{\vec{V}_J \vec{N}^T}{\vec{V}_{V\backslash \{i\}\cap J} \vec{N}^T}$.
\end{enumerate}

Applying the sign function does not introduce new dependencies that would alter the conditional independence structure inherent in the original GMRF. That is to say,
\begin{enumerate}
    \item For any $i, j$ not equal or adjacent, $\ConditionallyIndependent{\vec{V}_i \vec{N}^T}{\vec{V}_j \vec{N}^T}{\text{sign}(\vec{V}_{V\backslash \{i, j\}} \vec{N}^T)}$,
    \item For any $i\in V$ and $J \subset V$ not containing or adjacent to $i$, $\ConditionallyIndependent{\vec{V}_i \vec{N}^T}{\vec{V}_J \vec{N}^T}{\text{sign}(\vec{V}_{V\backslash \{i\}\cap J} \vec{N}^T)}$.
\end{enumerate}
Furthermore, since the sign function is symmetric on normal random variables with zero mean, the marginal probability distributions of $\text{sign}(\vec{V}_i \vec{N}^T)$ and $\text{sign}(\vec{V}_j \vec{N}^T)$ are also conditionally independent for any $i, j$ not equal or adjacent. For example,
\beq
\begin{aligned}
P(\text{sign}(\vec{V}_i \vec{N}^T)=1,\, &\text{sign}(\vec{V}_j \vec{N}^T)=1|\text{sign}(\vec{V}_{V\backslash \{i, j\}} \vec{N}^T))\\
=&P(\vec{V}_i \vec{N}^T > 0,\, \vec{V}_j \vec{N}^T > 0|\text{sign}(\vec{V}_{V\backslash \{i, j\}} \vec{N}^T))\\
=&P(\vec{V}_j \vec{N}^T > 0|\text{sign}(\vec{V}_{V\backslash \{i, j\}} \vec{N}^T))\, P(\vec{V}_j \vec{N}^T > 0|\text{sign}(\vec{V}_{V\backslash \{i, j\}} \vec{N}^T))\\
=&P(\text{sign}(\vec{V}_i \vec{N}^T)=1|\text{sign}(\vec{V}_{V\backslash \{i, j\}} \vec{N}^T))\, P(\text{sign}(\vec{V}_j \vec{N}^T)=1|\text{sign}(\vec{V}_{V\backslash \{i, j\}} \vec{N}^T)).
\end{aligned}
\eeq

Thus we have:
\begin{enumerate}
    \item For any $i, j$ not equal or adjacent, $\ConditionallyIndependent{U_i}{U_j}{U_{V\backslash \{i, j\}}}$,
    \item For any $i\in V$ and $J \subset V$ not containing or adjacent to $i$, $\ConditionallyIndependent{U_i}{U_J}{U_{V\backslash \{i\}\cap J}}$.
\end{enumerate}
So the $U_i$ also form a Markov random field respect to the same dependency graph $G$. We then define $\tilde{\lambda}$ as the maximum degree of $G$, which equals 2 times the number of nonzero $R(K)$s. 


Next, we can use Eq.~(\ref{varchi}) to write the variance of $\chi_U(T)$ as:
\[
\frac{4\pi^2}{\tau^4} \text{Var}(\chi_U(T))
= \sum_{ijkl} \Lambda_{ijkl} B_{ij}(S) B_{kl}(S),
\]
where we define
\[
\Lambda_{ijkl} = \EE(U_iU_jU_kU_l) - \EE(U_iU_j)\EE(U_kU_l).
\]
Note that $U_i^2 = 1$. Hence, whenever $i=j$ or $k=l$, we have $\Lambda_{ijkl} = 0$. Thus we can write:
\begin{equation}
\label{eqn-varchi-sum}
\frac{4\pi^2}{\tau^4} \text{Var}(\chi_U(T))
= \sum_{i\neq j,k\neq l} \Lambda_{ijkl} B_{ij}(S) B_{kl}(S).
\end{equation}

Now consider any $(i,j,k,l) \in \set{1,...,M}^4$ such that $i\neq j$ and $k\neq l$. Then $(i,j,k,l)$ must belong to one of the cases described below. For each of these cases, we will upper-bound the corresponding contribution to the sum in Eq.~(\ref{eqn-varchi-sum}):
\begin{itemize}
\item Case 1: $\set{i,j} = \set{k,l}$. Then we have the following upper bound:
\[
\begin{split}
\sum_{(i,j,k,l)\text{ in case 1}} &\Lambda_{ijkl} B_{ij}(S) B_{kl}(S) 
\leq 2\sum_{i\neq j} \abs{B_{ij}(S)}^2 
\leq 2M\sum_{\ell\neq 0} \abs{\hat{S}_\ell}^2,
\end{split}
\]
where we used the fact that $\Lambda_{ijij} = \Lambda_{ijji} = 1 - \EE(U_iU_j)^2 \leq 1$, and the definition of $B_{ij}(S)$ in Eq.~(\ref{bdef}).

\item Case 2: $\abs{\set{i,j} \cap \set{k,l}} = 1$. Without loss of generality, suppose that $j=k$. Then we have $\Lambda_{ijkl} = \Lambda_{ijjl} = \EE(U_i U_l) - \EE(U_i U_j) \EE(U_j U_l)$. Now let $H$ be the induced subgraph of $G$ on the set of vertices $\set{i,j,l}$. $H$ belongs to one of the following cases:
\begin{itemize}
    \item Case 2(a): $H$ has no edges, i.e., all three vertices in $H$ are isolated. This implies that $\Lambda_{ijjl} = 0$, and thus does not contribute to the above sum.
    
    \item Case 2(b): $H$ consists of two connected components. This implies that one of the vertices in $H$ is isolated (has no edges). If the isolated vertex is $i$ or $l$, then $\Lambda_{ijjl} = 0$, and thus does not contribute to the above sum. If the isolated vertex is $j$, then $\Lambda_{ijjl} = \EE(U_i U_l) \leq 1$, and $i$ and $l$ are connected by an edge in $H$. Then we have the following upper bound:
    \[
    \begin{split}
    \sum_{(i,j,j,l)\text{ in case 2(b)}} &\Lambda_{ijjl} B_{ij}(S) B_{jl}(S) \\
    &\leq \tfrac{1}{2} \sum_{(i,j,j,l)\text{ in case 2(b) s.t. }i\sim l\text{ in }H} \abs{B_{ij}(S)}^2 + \abs{B_{jl}(S)}^2 \\
    &\leq \tfrac{1}{2}\tilde{\lambda} \sum_{i\neq j} \abs{B_{ij}(S)}^2 + \tfrac{1}{2}\tilde{\lambda} \sum_{j\neq l} \abs{B_{jl}(S)}^2 
    \leq \tilde{\lambda}M \sum_{\ell\neq 0} \abs{\hat{S}_\ell}^2,
    \end{split}
    \]
    where we used the arithmetic-geometric mean inequality, and the fact that for any choice of $i$, there are at most $\tilde{\lambda}$ choices of $l$ that contribute to the sum, and vice versa.

    \item Case 2(c): $H$ consists of a single connected component. Note that $\Lambda_{ijjl} \leq 2$. Using a similar argument as in case 2(b), we have the following upper bound:
    \[
    \begin{split}
    \sum_{(i,j,j,l)\text{ in case 2(c)}} &\Lambda_{ijjl} B_{ij}(S) B_{jl}(S) \\
    &\leq \sum_{(i,j,j,l)\text{ in case 2(c) s.t. }H\text{ is connected}} \abs{B_{ij}(S)}^2 + \abs{B_{jl}(S)}^2 \\
    &\leq 2\tilde{\lambda} \sum_{i\neq j} \abs{B_{ij}(S)}^2 + 2\tilde{\lambda} \sum_{j\neq l} \abs{B_{jl}(S)}^2 
    \leq 4\tilde{\lambda}M \sum_{\ell\neq 0} \abs{\hat{S}_\ell}^2,
    \end{split}
    \]
    where we used the fact that for any choices of $i$ and $j$ (respectively $j$ and $l$), there are at most $2\tilde{\lambda}$ choices of $l$ (respectively $i$) that contribute to the sum.
\end{itemize}

\item Case 3: $\set{i,j} \cap \set{k,l} = \emptyset$. Let $H$ be the induced subgraph of $G$ on the set of vertices $\set{i,j,k,l}$. $H$ belongs to one of the following cases:
\begin{itemize}
    \item Case 3(a): There exists at least one vertex in $H$ has no edges, i.e., it is isolated (disconnected) from all of the other vertices. Without loss of generality, let this be vertex $i$. This implies that 
    \[
    \Lambda_{ijkl} = \EE(U_i)\EE(U_jU_kU_l) - \EE(U_i)\EE(U_j)\EE(U_kU_l) = 0, 
    \]
    and thus does not contribute to the above sum.
    
    \item Case 3(b): $H$ consists of two edges $(i,j)$ and $(k,l)$. This implies that 
    \[
    \Lambda_{ijkl} = \EE(U_iU_j)\EE(U_kU_l) - \EE(U_iU_j)\EE(U_kU_l) = 0, 
    \]
    and thus does not contribute to the above sum.
    
    \item Case 3(c): $H$ consists of two edges $(i,k)$ and $(j,l)$. Then we have the following upper-bound:
    \[
    \begin{split}
    \sum_{(i,j,k,l)\text{ in case 3(c)}} &\Lambda_{ijkl} B_{ij}(S) B_{kl}(S) \\
    &\leq \tfrac{1}{2} \sum_{(i,j,k,l)\text{ in case 3(c) s.t. }i\sim k,j\sim l\text{ in }H} \abs{B_{ij}(S)}^2 + \abs{B_{kl}(S)}^2 \\
    &\leq \tfrac{1}{2}\tilde{\lambda}^2 \sum_{i\neq j} \abs{B_{ij}(S)}^2 + \tfrac{1}{2}\tilde{\lambda}^2 \sum_{k\neq l} \abs{B_{kl}(S)}^2 
    \leq \tilde{\lambda}^2 M \sum_{\ell\neq 0} \abs{\hat{S}_\ell}^2,
    \end{split}
    \]
    where we used the fact that for any choices of $i$ and $j$ (respectively $k$ and $l$), there are at most $\tilde{\lambda}^2$ choices of $k$ and $l$ (respectively $i$ and $j$) that contribute to the sum.
    
    \item Case 3(d): $H$ consists of two edges $(i,l)$ and $(j,k)$. For this case, we have the same upper-bound as for case 3(c).

    \item Case 3(e): $H$ consists of a single connected component. Note that $\Lambda_{ijkl} \leq 2$. Then we have the following upper bound:
    \[
    \begin{split}
    \sum_{(i,j,k,l)\text{ in case 3(e)}} &\Lambda_{ijkl} B_{ij}(S) B_{jl}(S) \\
    &\leq \sum_{(i,j,k,l)\text{ in case 3(e) s.t. }H\text{ is connected}} \abs{B_{ij}(S)}^2 + \abs{B_{jl}(S)}^2 \\
    &\leq 6\tilde{\lambda}^2 \sum_{i\neq j} \abs{B_{ij}(S)}^2 + 6\tilde{\lambda}^2 \sum_{j\neq l} \abs{B_{jl}(S)}^2 
    \leq 12\tilde{\lambda}^2 M \sum_{\ell\neq 0} \abs{\hat{S}_\ell}^2,
    \end{split}
    \]
    where we used the fact that for any choices of $i$ and $j$ (respectively $k$ and $l$), there are at most $(2\tilde{\lambda}) (3\tilde{\lambda})$ choices of $k$ and $l$ (respectively $i$ and $j$) that contribute to the sum.
\end{itemize}
\end{itemize}

Combining the above results, and using Parseval's identity, we get the following upper bound on the variance of $\chi_U(T)$:
\beq
\begin{aligned}
\label{higherterm}
\Var(\chi_U(T))\leq \frac{(7\tilde{\lambda}^2 + 10\tilde{\lambda} + 1)M\tau^3}{\pi}||S(\omega)\sinc^2(\tfrac{\omega\tau}{2})-\langle S(\omega)\sinc^2(\tfrac{\omega\tau}{2})\rangle||^2_{L2}.
\end{aligned}
\eeq
Note that the bound in Eq.~(\ref{higherterm}) also holds for previous cases in Eq.~(\ref{chibound}) and Eq.~(\ref{chibound0}) when one sets $\tilde{\lambda}=0$ and $\tilde{\lambda}=1$, respectively.

\subsection{Variance of the decay exponent: a special case}
In this section, we illustrate a special case in Section \ref{sec-variance-decay-exponent-general-case} Case 3(e), where $H$ consists of a single connected component with $l=k+1=j+2=i+3$. We compute the exact value of the fourth order term $\EE(U_iU_jU_kU_l)$, which is rather complicated. 
We can again utilize the geometric interpretation of random pulses with hyper-plane tessellation: we now have 4 hyper-planes, indicated by 4 normal vectors, $\vec{V}_1=(a_0, a_1, 0, 0, 0)$, $\vec{V}_2=(0, a_0, a_1, 0, 0)$, $\vec{V}_3=(0, 0, a_0, a_1, 0)$ and $\vec{V}_4=(0, 0, 0, a_0, a_1)$, with $a_0a_1=\tfrac{\pi}{2}\sin(R(1))$. For simplicity, we remove the irrelevant zeros in the vectors to reduce the dimension. 

The hyper-planes cut the whole space into 16 pieces. In half of the separated spaces, the product $U_iU_jU_kU_l$ equals 1, while in the other half of the spaces, $U_iU_jU_kU_l$ equals $-1$. The problem now is to find the total solid angles of spaces with the same value. Define $\Omega_+$ to be the total solid angle with $U_iU_jU_kU_l=1$, and $\Omega_-$ for the part with $U_iU_jU_kU_l=-1$, so we have
\beq
\label{solidtoterm}
\EE(U_iU_jU_kU_l)=\frac{\Omega_+ -\Omega_-}{\Omega_+ +\Omega_-}.
\eeq

This can be calculated with the help of Ribando's formula \cite{ribando2006measuring}:

\beq
\label{ribando}
\Omega=\Omega_d \frac{|\text{det}(V)|}{(2\pi)^{d/2}}\sum_{\vec{a}}[\frac{(-2)^{\sum_{i<j}a_{ij}}}{\prod_{i<j}a_{ij}!}\prod_i \Gamma(\frac{1+\sum_{m\neq i}a_{im}}{2})]\vec{\alpha}^{\vec{a}}.
\eeq

In Ribando's formula, $\Omega_d$ represents the solid angle subtended by the $(d-1)$-dimensional spherical surface of a unit sphere, which equals $8\pi^2/3$ in our case ($d=5$). $V$ denote the matrix formed by the unit vectors defining the angle, which are the 4 normal vectors, $\vec{V}_i$. The multivariable $\vec{\alpha}$ contains the inner products of $\vec{V}_i$, with $\alpha_{ij}=\vec{V}_i\cdot \vec{V}_j$, $1\leq i \leq j \leq 4$. The vector $\vec{a}$ is the corresponding integer multiexponent, with $a_{ji}=a_{ij}$ and $\vec{\alpha}^{\vec{a}}=\prod \alpha_{ij}^{a_ij}$. Define matrix $M$ whose diagonal elements are $1$ and off diagonal elements equals $-|\alpha_{ij}|$. Equation (\ref{ribando}) will  converge to the solid angle defined by the vectors if and only if $M$ is positive definite, which is true in our scenario.

Note that all 16 solid angles can be defined by the 4 normal vectors by adding negative signs to them. So the multivariables that determine the solid angles can be written as $\vec{\alpha}=(\pm \tfrac{\pi}{2}\sin(R(1)), 0, 0, \pm \tfrac{\pi}{2}\sin(R(1)),0, \pm \tfrac{\pi}{2}\sin(R(1)))$. The 16 pieces are reduced to 8 due to the symmetry of the hyper-plane tessellation. Furthermore, we define $\Omega_{\pm \pm \pm}$ to be the specific solid angle whose notes indicates the signs of the three non-zero elements in $\vec{\alpha}$, we have
\beq
\begin{aligned}
\label{solidangles}
\Omega_- &= 2(\Omega_{+++}+\Omega_{---}+\Omega_{-+-}+\Omega_{+-+}),\\
\Omega_+ &= 2(\Omega_{-++}+\Omega_{--+}+\Omega_{+--}+\Omega_{++-}).
\end{aligned}
\eeq

Now, substituting Eq.~(\ref{solidangles}) and Eq.~(\ref{ribando}) into Eq.~(\ref{solidtoterm}), we can find the value of $\EE_{\{i,j,k,l\}\in F}(U_iU_jU_kU_l)$ as an infinite series involving $R(1)$:
\beq
\begin{aligned}
    \EE(U_iU_jU_kU_l)
    =&\frac{\sum_{a_1a_2a_3}G(a_1,a_2,a_3,R(1)){(1+(-1)^{a_2}-(-1)^{a_1}-(-1)^{a_3})}}{\sum_{a_1a_2a_3}G(a_1,a_2,a_3,R(1)){(1+(-1)^{a_2}+(-1)^{a_1}+(-1)^{a_3})}},\\
    G(a_1,a_2,a_3,R(1)):=&\frac{(-2)^{a_1+a_2+a_3}}{\prod_i a_1!a_2!a_3!}\Gamma(\frac{1+a_1}{2})\Gamma(\frac{1+a_1+a_2}{2})\Gamma(\frac{1+a_2+a_3}{2})\Gamma(\frac{1+a_3}{2})\\
    &\cdot(\tfrac{\pi}{2}\sin(R(1)))^{a_1+a_2+a_3}(1+(-1)^{a_1+a_2+a_3}).
\end{aligned}
\eeq
One trivial example is that when $R(1)=0$, $\EE(U_iU_jU_kU_l)=0$. And when $R(1)=1/3$, $\EE(U_iU_jU_kU_l)=2/15$.

\subsection{Variance of the decay exponent: the case of the alternative generator}
Another special case is the alternative random pulse generator, described in Section \ref{sec-alternative-generator}. It can be considered as simplified version of the general one, where the dependency graph $G$ is of degree $1$, i.e., $\tilde{\lambda} = 1$. Similar to the general model, we have
\begin{equation}
\frac{4\pi^2}{\tau^4} \text{Var}(\chi_U(T))
= \sum_{i\neq j,k\neq l} \Lambda_{ijkl} B_{ij}(S) B_{kl}(S).
\end{equation}
The remaining cases of $(i, j, k, l)$  that have a nonzero contribution to the sum are: Case 1, Case 2(b), Case 3(c) and Case 3(d). It can be calculated that
\beq
    \label{chibound0}
    \begin{aligned}
        \text{Var}(\chi_U(T)) \leq \frac{4M\tau^3}{\pi}||S(\omega)\sinc^2(\tfrac{\omega\tau}{2})-\langle S(\omega)\sinc^2(\tfrac{\omega\tau}{2})\rangle||^2_{L2}.
    \end{aligned}
\eeq



\subsection{Experimental estimation of the decay exponent}
\label{secd3}
In order to estimate the $\chi_U(T)$ accurately, we need to set the parameters $\tau$ and $M$ appropriately, perform multiple runs of the experiment (using different pulse sequences), and average the results. Here, we describe how to do this.

First, we choose the period $\tau$ of the random pulse sequences. We assume that we are given some high-frequency cutoff $\omega_c$, such that $S(\omega)$ is supported inside the interval $[-\omega_c, \omega_c]$.

Next, we choose the length $M$ of the random pulse sequences. Our goal is to ensure that $\chi_U(T)$ is of order 1, so that it can be accurately estimated by observing events that occur with probability $P_0(\vec{U}) = \tfrac{1}{2}(1 + e^{-\chi_U(T)})$. Based on equation (\ref{ews}), we know that $\chi_U(T)$ is linearly dependent on $M$. Here, we assume that we are given an initial guess for $\chi_U(T)$ (call it $\chi_g$) that has the correct order of magnitude. We then use $\chi_g$ to choose the value of $M$. 

To be concrete, let us assume that $\chi_g$ lies within the range 
\begin{equation}
\frac{\chi_U(T)}{10} \leq \chi_g \leq 10 \chi_U(T).
\end{equation}
Then we set 
\begin{equation}
M = \biggl\lceil \frac{2\pi}{\tau^2 \chi_g} \biggr\rceil.
\end{equation}
The actual value of $\chi_U(T)$ will fluctuate around its expectation, but these fluctuations will less than 1 with high probability, as shown later by equation (\ref{eqn-Q-not-too-large}).

Let us fix two parameters $N_1$ and $N_2$. We then choose $N_1$ random pulse sequences (let us call them $\vec{U}^{(j)}$, for $j = 1,\ldots,N_1$). For each pulse sequence $\vec{U}^{(j)}$, we run $N_2$ trials of the experiment, and obtain measurement outcomes 
\begin{equation}\label{eqn-est-Xjk}
X_{jk} = \begin{cases}
 1 &\text{ if we observe $\ket{0}$} \\
-1 &\text{ if we observe $\ket{1}$},
\end{cases}
\end{equation}
for $k = 1,\ldots,N_2$. Note that $X_{jk}$ has expectation value $\EE(X_{jk} \,|\, \vec{U}^{(j)}) = \exp(-\chi_{U_j}(T))$. This motivates us to define an estimator $\chi$ as follows: 
\begin{equation}\label{eqn-est-Yj}
Y_j = \frac{1}{N_2} \sum_{k=1}^{N_2} X_{jk},
\end{equation}
\begin{equation}\label{eqn-est-Zj}
Z_j = -\log Y_j,
\end{equation}
\begin{equation}\label{eqn-est-chiexp}
\chi^{\text{exp}} = \frac{1}{N_1} \sum_{j=1}^{N_1} Z_j.
\end{equation}

\subsection{Accuracy of the estimator}
\label{sec-acc}
At last, we can combine the previous results and derive a bound for $\chi^\text{exp}$. Recall that the variance of $\chi_U$ is bounded by Eq. (\ref{higherterm}). Using Chebyshev's  inequality  plus a union bound, we can prove that  for all $j$ over $\vec{U}^{(j)}$,
\begin{equation}\label{eqn-Q-not-too-large}
P(|\chi_{U_j}(T)-\EE(\chi)| > N_1 \sqrt{\frac{(7\tilde{\lambda}^2 + 10\tilde{\lambda} + 1)M\tau^3}{\pi}}{||S(\omega)\sinc^2(\tfrac{\omega\tau}{2})-\langle S(\omega)\sinc^2(\tfrac{\omega\tau}{2})\rangle||_{L2}}) \leq \frac{1}{N_1}. 
\end{equation}
Since $\EE(\chi)$ scales linearly with $M$, $\chi_{U_j}(T)$ is quite concentrated when $M$ is large. Following a Bernstein-type inequality for sub-exponential random variables \cite{vershynin2010introduction}, we obtain 
\begin{equation}\label{eqn-Q-to-S0}
\frac{1}{N_1} \sum_{j=1}^{N_1} \chi_{U_j}(T) \approx \EE(\chi_U(T)) \pm \sqrt{\frac{(7\tilde{\lambda}^2 + 10\tilde{\lambda} + 1)M\tau^3}{\pi}}\frac{||S(\omega)\sinc^2(\tfrac{\omega\tau}{2})-\langle S(\omega)\sinc^2(\tfrac{\omega\tau}{2})\rangle||_{L2}}{\sqrt{N_1}}.
\end{equation}

Conditioned on choices of the $\vec{U}^{(j)}$ that satisfy the above properties, we will now consider the probabilities over $X_{jk}$, which are independent Bernoulli random variables. According to Hoeffding's inequality, 
\beq
P(|Y_j-\exp(-\chi_{U_j}(T))|\geq t)\leq 2\exp(-\frac{N_2t^2}{2}).
\eeq
So $Y_j$ has a binomial distribution with exponentially-decaying tails, centered around $\exp(-\chi_{U_j}(T))$, with standard deviation $O(\exp(-\chi_{U_j}(T))/\sqrt{N_2})$. The random variable $Z_j$ takes the negative logarithm of $Y_j$, which also has exponentially-decaying tails centered around $\chi_{U_j}(T)$ and with standard deviation $O(\chi_{U_j}(T)/\sqrt{N_2})$. This holds because we are in the regime where the log function does not blow up too badly, because of (\ref{eqn-Q-not-too-large}). Then with high probability, 
\begin{equation}\label{eqn-S0hat-to-Q}
\chi^\text{exp} \approx \frac{1}{N_1} \sum_{j=1}^{N_1}  \chi_{U_j}(T) \pm \frac{O(\chi_{U_j}(T))}{\sqrt{N_1 N_2}}. 
\end{equation}
This follows from a Bernstein-type inequality for sub-exponential random variables \cite{vershynin2010introduction}. 

Now combining (\ref{eqn-Q-to-S0}) and (\ref{eqn-S0hat-to-Q}), we can see that 
\begin{equation}
\chi^\text{exp} \approx \EE(\chi) \pm \sqrt{\frac{(7\tilde{\lambda}^2 + 10\tilde{\lambda} + 1)M\tau^3}{\pi}}\frac{||S(\omega)\sinc^2(\tfrac{\omega\tau}{2})-\langle S(\omega)\sinc^2(\tfrac{\omega\tau}{2})\rangle||_{L2}}{\sqrt{N_1}} \pm O(\frac{\EE(\chi)}{\sqrt{N_1N_2}}).
\end{equation}
Note that there are two error terms. Increasing $N_1$ causes both error terms to decrease,  increasing $N_2$ only helps with the second error term. Thus the best strategy is to make $N_1$ large, and keep $N_2$ as small as possible. 

But how small can we choose $N_2$? If $N_2$ is too small, we have problems: $Y_j$ can be close to zero, which causes the log function to blow up when computing $Z_j$. To avoid these problems, we need to make sure that $1/\sqrt{N_2} \leq O(\exp(-\chi_{U_j}))$, or equivalently, $N_2 \geq \Omega(\exp(2\chi_{U_j}))$. 

Although we don't know the value of $\chi_{U_j}$ a priori, we can set $N_2$ adaptively to get close to this bound. For most choices of $\vec{U}^{(j)}$, $\chi_{U_j}$ is at most some constant, hence setting $N_2$ to be a constant should be good enough. However, since we are using $N_1$ random pulse sequences ($\vec{U}^{(j)}$, for $j = 1,\ldots,N_1$), we expect there could be a few $\vec{U}^{(j)}$ where $\chi_{U_j}$ could be as large as equation (\ref{eqn-Q-not-too-large}). For those $\vec{U}^{(j)}$, we could obtain a large estimation for $\chi_{U_j}$ and increase $N_2$ accordingly.

Another way to get rid of the problem with large $\chi_{U_j}$ is to use the median of means method \cite{jerrum1986random,nemirovskij1983problem}. We can construct $K$ independent sets of size $N_1$. For each individual set, we do the same experiment as we discussed before and get an estimation for the decay rate. We then find the median of their values. It is  guaranteed that for $K$ independent sets of size $N_1=34Z/\epsilon^2$ suffice to construct a median of mean estimator $\chi^\text{exp}_\text{median}$ that obeys
\beq
\label{median2}
\begin{aligned}
&\text{Pr}(|\chi^\text{exp}_\text{median}-\mathbb{E}(X)|\geq \epsilon) \leq 2e^{-K/2},
\end{aligned}
\eeq
where $Z=\tfrac{(7\tilde{\lambda}^2 + 10\tilde{\lambda} + 1)M\tau^3}{\pi}||S(\omega)\sinc^2(\tfrac{\omega\tau}{2})-\langle S(\omega)\sinc^2(\tfrac{\omega\tau}{2})\rangle||^2_{L2}$.

Finally, it is interesting to consider the most extreme case where we set $N_2 = 1$. This can be convenient in certain experimental setups, where it is possible to generate a fresh random pulse sequence during every run of the experiment, and where it is difficult to store and repeat a pulse sequence. In this case, the estimator $\chi^\text{exp}$ in Eqs.~(\ref{eqn-est-Xjk})-(\ref{eqn-est-chiexp}) does not work, because the log function diverges, as discussed above. However, one can still obtain crude estimates of the decay exponent $\chi$, using the following estimator:
\begin{equation}
\chi^\text{crude} = 1-X, \text{ where } X = \frac{1}{N_1} \sum_{j=1}^{N_1} X_{j1}.
\end{equation}

This estimator $\chi^\text{crude}$ is based on the following intuition. First, one computes $X$, which has expectation value $\EE(X) = \EE(\exp(-\chi_U(T)))$, where one averages over the choice of the random pulse sequence $\vec{U}$. Then one estimates $\chi_U(T)$, using the approximation $\exp(-\chi_U(T)) \approx 1-\chi_U(T)$, which is valid when $\chi_U(T) \ll 1$. 

Formally, the estimator $\chi^\text{crude}$ satisfies the following upper bound, which relates $\EE(\chi^\text{crude})$ and $\EE(\chi_U(T))$:
\begin{equation}\label{eqn-crude-upper}
\EE(\chi^\text{crude}) \leq \EE(\chi_U(T)).
\end{equation}
This follows from the inequality $1-e^{-x} \leq x$. 

In addition, $\chi^\text{crude}$ obeys a weak lower bound, which is useful when $\EE(\chi^\text{crude})$ is small. Essentially, this upper bounds the probability that $\EE(\chi^\text{crude})$ will be much smaller than the true value of $\chi_U(T)$: 
\begin{equation}\label{eqn-crude-lower}
\Pr(\chi_U(T) \geq \lambda \EE(\chi^\text{crude})) \leq \EE(\chi^\text{crude}) + \frac{1}{\lambda} \quad (\forall \lambda \geq 1).
\end{equation}
To prove this, let $\delta = \EE(\chi^\text{crude})$. Then we have 
\begin{equation}
\begin{split}
1-\delta = \EE(\exp(-\chi_U(T))) &\leq \Pr(\chi_U(T)\geq\lambda\delta) e^{-\lambda\delta} + \Pr(\chi_U(T)<\lambda\delta) \\
&= 1 - \Pr(\chi_U(T)\geq\lambda\delta) (1-e^{-\lambda\delta}).
\end{split}
\end{equation}
After some algebra, we get
\begin{equation}
\begin{split}
\Pr(\chi_U(T)\geq\lambda\delta) &\leq \frac{\delta}{1-e^{-\lambda\delta}}
= \delta + \frac{\delta e^{-\lambda\delta}}{1-e^{-\lambda\delta}}
= \delta + \frac{\delta}{e^{\lambda\delta}-1} \\
&\leq \delta + \frac{1}{\lambda}.
\end{split}
\end{equation}


\section{Compressed sensing}

\subsection{Random Sparse Spectra}
\label{sec-random-sparse-spectra}
In the main paper, we generate random sparse spectra to further examine the precision of CS (see Fig.~\ref{cs}). 
These random spectra are generated in the following way. We initialize a vector of dimension $N$, with every coordinate equal to zero. Then we randomly pick $s$ coordinates, where $s$ is the prescribed sparsity. Next, we set the amplitudes of those chosen coordinates to be uniformly random in $[0,1]$. Finally, we normalize the vector so that its $\ell_1$ norm equals $s$. The average value of the peaks remains the same regardless of $s$, thus we can compare the accuracy of the CS method under different random spectra and different sparsities.

\subsection{LASSO}
\label{sec-lasso}
\begin{figure}
\includegraphics[width=0.4\textwidth]{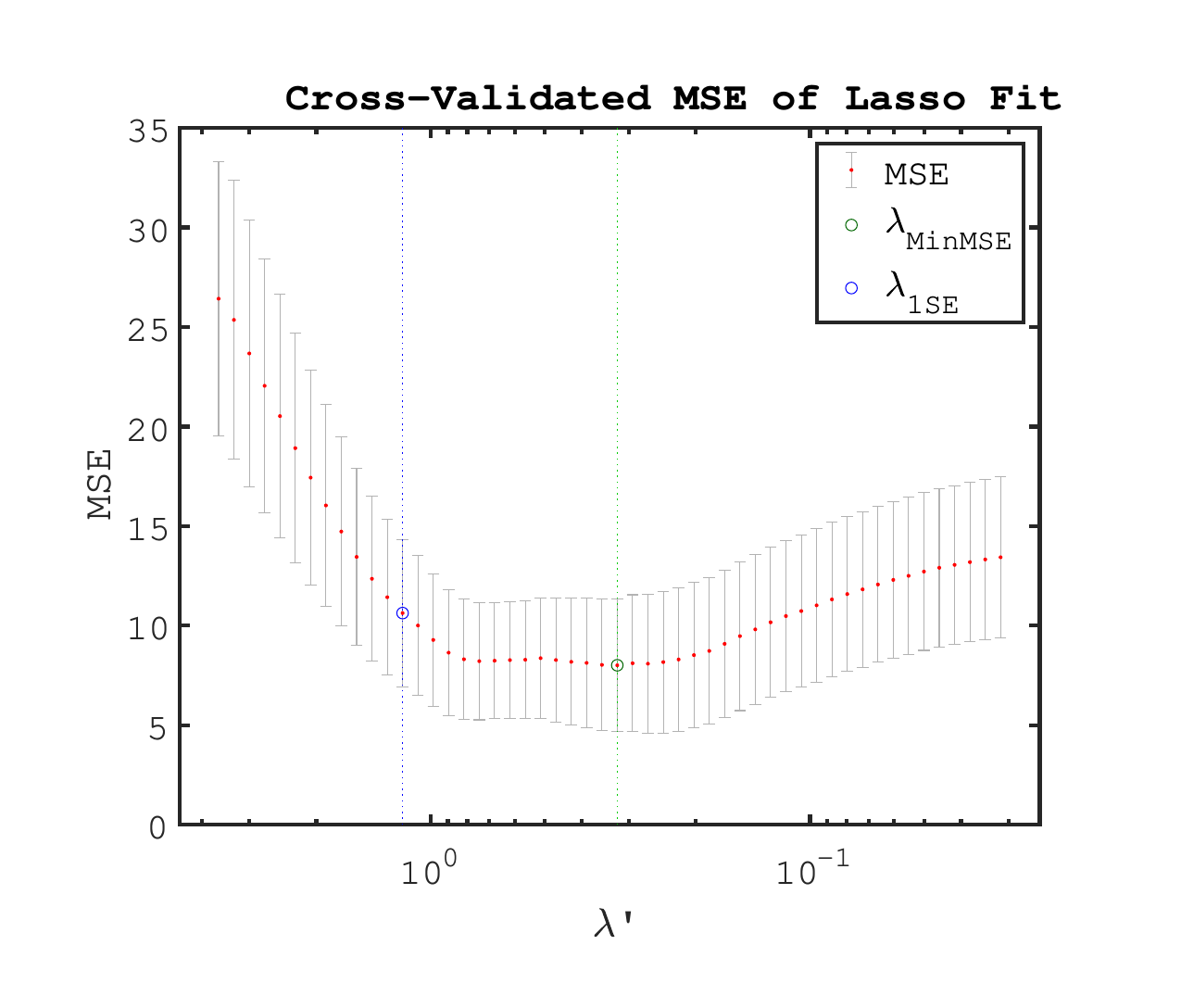}
\caption{We use the 10-fold cross validation method to find out the best value of $\lambda'$. The green circle and dotted line locate the $\lambda'$ with the minimum cross-validated mean squared error ($\lambda_{MinMSE}$). The blue circle and dotted line locate the sparsest model within one standard error of the minimum MSE ($\lambda_{1SE}$).}
\label{fig:FIG6ba}
\end{figure}

In real physical systems, the noise spectra are usually not ideally sparse. We may adopt suitable data analysis techniques to better process compressed sensing. For example, we can apply the LASSO (east absolute shrinkage
and selection) algorithm, which solves the following problem,
\begin{equation}
\label{lassoeq}
    \begin{aligned}
    \text{find $S^* \in \RR^N$ that minimize}\\
    \tfrac{1}{2}||y^T-AS^{*T}||_{L2}^2+\lambda||S^*||_{L1}.
    \end{aligned}
\end{equation}
This can be viewed as a Lagrangian relaxation of Eq.~(\ref{CSconvec-0}), which is useful when one is given data $y$ that contain noise or errors. 
The coefficient $\lambda$ controls the relative strength  between the two parts in Eq.~(\ref{lassoeq}). In general, when the noise is strong, one need a larger $\lambda$ to ensure the $L_1$ regularization term still matters. Meanwhile, $\lambda$ cannot be too large; otherwise, the result will be too biased. Here, we use the cross-validation (CV) method to establish the value of $\lambda$. As is shown in an illustration in Fig. \ref{fig:FIG6ba}, we generate series of $\lambda$ and do a cross-validation.  The green circle and dotted line locate the $\lambda'$ with the minimum cross-validated mean squared error ($\lambda_{MinMSE}$). The blue circle and dotted line locate the sparsest model within one standard error of the minimum MSE ($\lambda_{1SE}$). In this work we choose $\lambda_{1SE}$ as our result because our goal is to separate the major peaks. $\lambda_{1SE}$ can help us figure out the isolated peaks while remove other redundant information.

\end{document}